\documentclass[a4paper,11pt]{article}

\usepackage{jcappub} % for details on the use of the package, please
\usepackage[T1]{fontenc} % if needed
\usepackage{amsthm,amsmath,amssymb}
\usepackage{mathrsfs}
\usepackage{graphicx}
\usepackage{multirow}

\title{Geometric Structure of Multi-Form-Field Isotropic Inflation and Primordial Fluctuations }

\author{Chong-Bin Chen,}
\author{Jiro Soda}

\affiliation[a]{Department of Physics, Kobe University, Kobe 657-8501, Japan}

\emailAdd{chongbin@stu.kobe-u.ac.jp}
\emailAdd{jiro@phys.sci.kobe-u.ac.jp}

\abstract{
An inflationary scenario is expected to be embedded into an ultraviolet (UV)
complete theory such as string theory. The effect of UV complete theories may appear as nontrivial kinetic terms in the low energy effective field theory, which provides a nontrivial
geometry in field space. In this paper, we study the effect of the geometry
of multi-form-field space on an inflationary scenario. 
In particular, we focus on the geometric destabilization mechanism
which induces the phase transition from the conventional slow-roll inflation
to a novel inflationary scenario. Anisotropic inflation is a typical example of the new phase. To conform to observations, we restrict us to isotropic configuration of form fields.
 We clarify the conditions for the onset of the destabilization and
 reveal the geometric structure of attractors after the destabilization.
 We classify the viable models from the observational point of view. 
 We also investigate the features of the primordial fluctuations and 
 find the similarity to hyperbolic inflation. 
 By calculating the power spectrum, we make several
 phenomenological predictions which are useful to discriminate our models from others inflation models. We found the scalar-to-tensor $r$ will be suppressed by large one-form gauge fields, while it has the same order as the slow roll parameter $r\sim\mathcal{O}(1)\epsilon$ for large two-from gauge fields.}

\begin{document}
\maketitle
\flushbottom

\section{Introduction}
An inflationary scenario provides a solution to the  flatness and horizon problems \cite{Guth:1980zm,Sato:1981ds,Linde:1981mu,Albrecht:1982wi}. Remarkably, single-field inflation models account for the primordial fluctuations of the large scale structure of the universe and have been excellently tested by the latest observations \cite{Planck:2018jri,Array:2015xqh}. However,  inflation models must be embedded into an ultraviolet (UV) complete theory, e.g. string theory ~\cite{Vafa:2005ui,Ooguri:2006in,Obied:2018sgi,Agrawal:2018own,Garg:2018reu},
as a low energy effective field theory. 
In general, there are many (quasi-)heavy fields in the UV complete theory which should be taken into account when we discuss inflation. 
One can integrate them out and describe the system by an effective single-field theory with some corrections, for example, the modification to the effective speed of sound which can be less than the speed of light \cite{Cheung:2007st,Achucarro:2010da,Achucarro:2012sm}.
Thus, in contrast to the single-field case, taking into account these new heavy fields may result in some new features in the evolution of the system, both in the background and perturbations. 
%If the background solution of a multi-scalar-field inflationary system follows a non-geodesic trajectory in the field space, the adiabatic perturbations will be coupled to the entropic perturbations hence power spectrum can be significantly changed on large scales ~\cite{Gordon:2000hv,Wands:2000dp}. Nevertheless, since the single-field inflation well explains observations, such non-inflationary degrees of freedom should be very heavy or becomes irrelevant higher dimensional operators in the single-field inflationary model on inflationary energy scales. The importance of these couplings with new degrees of freedom arises through the renormalization-group flow to a UV energy scale of the inflation. In other words, 
%The heavy scalar degrees of freedom play a role of "spectator" and are suppressed by the large mass of the heavy scalar field during the inflation. 

The additional degrees of freedom in the inflationary models motivated from the string theory and supergravity may provide a non-canonical kinetic term of the scalar fields. For example, if we consider the effective field theory containing an operator $-(\partial\phi_1)^2\phi_2^2/M^2$, where the inflaton $\phi_1$ is coupled to an extra scalar field $\phi_2$ at energy scale $M$. This equivalently modify the kinetic term of scalar fields to $(1+2\phi_2^2/M^2)(\partial\phi_1)^2+(\partial\phi_2)^2$, where the metric of the field space is not flat anymore. Thus, in the contexts of beyond the Standard Model, it seems natural to consider
a general form of the kinetic terms
\begin{equation}
    S_{\text{kin}}=-\frac{1}{2}
    \int d^4x\sqrt{-{\tilde g}}
    c_{ab} (\phi)\partial^\mu\phi^a \partial_\mu\phi^{b} \ ,
    \qquad \phi^a=(\phi^1,\phi^2)
    \ .\label{scalarkin}
\end{equation}
where ${\tilde g}$ is the determinant of a spacetime metric ${\tilde g}_{\mu\nu}$ and $c_{ab}$ is the metric of the field space. Recently,   it is shown that a geometric 
destabilization occurs for the non-flat geometry of 
scalar field space (\ref{scalarkin})~\cite{Renaux-Petel:2015mga,Garcia-Saenz:2018ifx,Renaux-Petel:2017dia,Cicoli:2018ccr,Cicoli:2019ulk}. 
Specifically, for a two-scalar-field inflation one can 
deduce the equation of motion for the entropic perturbations $Q_s$ on large scales and
the mass squared term is given by
\begin{equation}
   M_s^2\equiv V_{NN}-H^2\eta_{N}^2+\epsilon_H H^2R_{\text{fs}}M_{\text{pl}}^2
   \ ,\label{sms}
\end{equation}
where $M_{\text{pl}}$ is the Planck mass, $V_{NN}$ is the mass term in the potential, $H$ is the Hubble parameter, $R_{\text{fs}}$ is the curvature scale of the field space, $\eta_N$ is the turn rate of the trajectory in the field space, and $\epsilon_H$ is the slow roll parameter.  The third term is the contribution from the effect of the geometry of the field space which is not necessarily small. If we assume $\phi_1$ is the inflaton and $\phi_2$ is a very heavy  extra field ($m_2\gg H$) so that it stays at the bottom of its potential, the system is in the conventional slow roll inflation phase driven by $\phi_1$.  Hence the second term in (\ref{sms}) vanishes because of $\eta_N=0$. If the field space has a large negative curvature $R_{\text{fs}}<0$, the mass squared would be tachyonic owing to the third term in (\ref{sms}), i.e., 
\begin{equation}
M_s^2|_{\eta_N=0}<0. 
\end{equation}
This is akin to the mechanism of hybrid inflation \cite{Linde:1993cn}, which includes a second "waterfall" field apart from the inflaton. This waterfall field becomes tychyonic 
and the conventional inflation phase ends, owing to the exponentially growth of the perturbation. 

In general, after the destabilization, the system will be settled into a second inflationary phase, where the heavy scalar field $\phi_2$ is excited and leaves away from the bottom of the potential. In the field space, the trajectory of scalar fields has non-zero turn rate $\eta_N\neq 0$ hence deviates from the original slow-roll trajectory along $\phi_1$. Recently, the inflationary attractors with such non-geodesic motion of scalar fields were studied. For example, the sidetracked inflation \cite{Garcia-Saenz:2018ifx}, which can be described by an effective single-field theory with an imaginary speed of sound. Furthermore, the geometric destabilization also occurs for the ultra-light extra field, i.e., $m_2\rightarrow 0$ \cite{Cicoli:2018ccr,Cicoli:2019ulk}. A prototypical attractor of such case is the hyperbolic inflation \cite{Brown:2017osf,Mizuno:2017idt,Bjorkmo:2019aev}, where the field space is a hyperbolic plane and the massless extra field $\phi_2$ can be regarded as an angular field rotating around the bottom of potential. Actually the hyperbolic inflation can be classified as one special case of the sidetracked inflation \cite{Christodoulidis:2019mkj}. All of these models enjoy the same features that the extra field rapidly turns~\cite{Bjorkmo:2019fls,Aragam:2020uqi}. That is, the energy of angular motion of the extra field is very large compared to the kinetic energy of inflaton. Hence the slow roll is allowed. Phenomenologically, these models can be distinguished from the single-filed one. The imaginary speed of sound results in a transient instability of fluctuations so that the perturbation modes experience a tachyonic growth before  the horizon crossing. At the same time, the tensor-to-scalar ratio is exponentially suppressed because of this instability. The bispectrum for this strongly non-geodesic motion 
is not exponentially enhanced. However, the shape of bispectrum has flattened configurations, which is quite different from the usual equilateral one in the single-field case \cite{Garcia-Saenz:2018vqf,Fumagalli:2019noh}.

We learned that the geometric structure 
of the multi-scalar field space can destabilize the conventional slow roll inflation and causes many interesting phenomenon. 
In this line of thought, the extra scalar fields can be replaced by gauge fields $A_\mu$. Indeed, the destabilization due to a gauge kinetic term occurs in the scalar-gauge-field models~\cite{Watanabe:2009ct}. For the Maxwell theory which is conformally coupled to gravity, the expansion of universe can be eliminated from the action. Hence the $U(1)$ gauge fields do not feel the expansion of the universe. To make the gauge fields relevant to inflation, One approach is to consider inflation driven by gauge fields involving $F^4$ terms \cite{Maleknejad:2011sq,Maleknejad:2011jw}.  Another one is to introduce the scalar fields coupling to the kinetic term of gauge fields. The later kind of models have been used to discuss the origin of magnetic fields on large scale in our universe \cite{Martin:2007ue,Demozzi:2009fu,Kanno:2009ei,Emami:2009vd,Talebian:2020drj,Talebian:2021dfq,Subramanian:2009fu,Fujita:2015iga,Fujita:2016qab,Fujita:2019pmi,Subramanian:2015lua}. Moreover, the 
models are called anisotropic inflationary models \cite{Watanabe:2009ct,Soda:2012zm,Watanabe:2010fh,Emami:2010rm,Yamamoto:2012tq,Ito:2016aai,Maleknejad:2012fw} described by the action
\begin{equation}
    S_{\text{kin}}=-\frac{1}{4}\int d^4x\sqrt{-{\tilde g}} f^2(\phi)F_{\mu\nu}F^{\mu\nu},
\end{equation}
where $F_{\mu\nu} = \partial_\mu A_\nu -\partial_\nu A_\mu$ is the field strength and the gauge kinetic function $f(\phi)=\exp{(2c\int V/V_{\phi}d\phi})$ with a parameter $c>1$ is chosen so that the gauge fields survive during inflation. The gauge field induces a second inflationary stage where the statistical anisotropy in the power spectrum of curvature fluctuations and other phenomenological signs are produced \cite{Soda:2012zm,Watanabe:2010fh,Ito:2016aai,Emami:2011yi,Emami:2013bk,Chen:2014eua,Emami:2015uva}. 
In this paper, we will show that the process can be interpreted as the geometric destabilization. Here the gauge kinetic function $f(\phi)$ can be regarded as the metric in the field space. We see the similarity between anisotropic inflation and  the hyperbolic inflation. 

Note that gauge fields can be regarded as one-form fields. We can also consider two-form gauge fields $B^a_{\mu\nu}$ whose field strength is $H^a_{\ \mu\nu\rho} = \partial_\mu B^a_{\nu\rho}+\partial_\nu B^a_{\rho\mu}+\partial_\rho B^a_{\mu\nu}$~\cite{Ohashi:2013mka,Ito:2015sxj,Ohashi:2013qba}.
Here, we are considering multi fields labelled by indices $a$.
The general kinetic terms of the model can be written as
\begin{equation}\label{actionall}
    S_{\text{kin}}=
    \int d^4x\sqrt{-{\tilde g}} \left[
    -\frac{1}{2}c_{ab} (\phi)\partial^\mu\phi^a \partial_\mu\phi^{b}
    -\frac{1}{4}f_{ab}(\phi)F^a_{\ \mu\nu}F^{b\mu\nu}
    -\frac{1}{4}g_{ab}(\phi)H^a_{\ \mu\nu\rho}H^{b \mu\nu\rho} \right]\ ,
\end{equation}
where $f_{ab}$ and $g_{ab}$ are the metrics of the field space of one-form and two-form gauge fields, respectively. Here, both the one-form and two-form gauge fields are massless, as in the hyperbolic inflation. Now the metric of the whole field space can be written by
\begin{equation}\label{G_metric}
G_{ab}(\phi)=
\left(\begin{array}{ccc}
    c_{ab}(\phi) & & \\
      & f_{ab}(\phi) & \\
      & & g_{ab}(\phi)
\end{array}\right).
\end{equation}
The exact power-law solutions for the hyperbolic inflation with a gauge field  have been studied previously~\cite{Chen:2021nkf,Do:2021pqk}. It is intriguing to study the above general models in detail. 

To study how gauge fields destabilize the conventional slow roll inflationary solutions, 
we start with models including only one scalar field ($c_{ab}=1$) and a triplet of isotropic multi-gauge fields.  Although some specific models have been discussed in previous studies, the connection between the geometric destabilization and the cosmological perturbations has not been explored. In particular, like as the cases of multi-scalar field (e.g. hyperbolic inflation), we consider the situation significant amount of gauge fields contribute to inflation dynamics, which causes the exponential growth of curvature perturbations inside the horizon. 
We consider the one-form gauge fields and the two-form gauge fields, separately. The isotropic configuration of gauge fields allows us to pick up the scalar parts of the perturbations and see how they can impact on the slow roll inflation.
Our main objectives in this paper are:
\begin{itemize}
\item[$\bullet$] Using the helicity decomposition, we pick up the scalar perturbation of gauge fields and derive the equations of motion of these perturbations. Then clarifying the conditions that these perturbation become tachyonic due to the geometrical effects.
\item[$\bullet$] As a consequence of geometrical destabilization, 
there are new attractor solutions. We classify solutions
that produce scale invariant power spectrum of gauge-field perturbations for a general field space metric $f_{ab}(\phi)$ or $g_{ab}(\phi)$.

\item[$\bullet$] At the linear perturbation level, we need to study the transient instability of perturbations before the horizon crossing. This instability results in an exponential growth of the perturbations. We analyze it numerically for one-form and two-form gauge fields, respectively.

\item[$\bullet$] We also calculate the power spectrum of curvature perturbations by taking into account the modes of scalar fields and gauge fields. We calculate the spectral index $n_s-1$ and the tensor-to-scalar ratio $r$ for these two models.
\end{itemize}

\begin{table}[]
\begin{center}
\renewcommand\arraystretch{1.7}
 \setlength{\tabcolsep}{2.1mm}{
\begin{tabular}{c|c|c|c|c|c}
\hline
Fields   & FS metric & BG field                          & SR variables                & Scalar fluc.                          & Tensor fluc.               \\ \hline
Scalar   & $c=1$     & $\phi$                            & $\epsilon_{\phi}$, $\eta$ & $\Delta_{\phi}=a\delta\phi$            & $\times$                   \\
One-form & $f$       & $A^a_{\ i}=\mathbb{A}\delta_{ai}$ & $\epsilon_E$, $h$, $\eta_h$ & $\Delta_Q=a\delta Q$, $\Delta_U=aU_i$ & $t_{ij}=2f\mathbb{T}_{ij}$ \\
Two-form & $g$       & $B^a_{\ k}=\mathbb{B}\delta_{ak}$ & $\epsilon_B$, $l$, $\eta_l$ & $\Delta_P=a\delta P$                  & $\times$                   \\
Gravity  & $\times$  & $\tilde{g}$, $q$, $\mathcal{N}$, $N^i$ & $\epsilon_H$                & $A$, $B$                              & $h_{ij}=aw_{ij}$           \\ \hline
\end{tabular}
}\caption{We present here some symbols that appear frequently in this paper. FS: Field Space. BG: Background. SL: Slow Roll. fluc.: fluctuation. $\times$: do not exist. We ignore the vector fluctuations because they do not contribute to the observations in our models. The definitions of SR variables can be found in Section \ref{Goa}. $\Delta_i$ and $t_{ij}$, $h_{ij}$ are canonical variables used in Section \ref{perturbation}.}
\end{center}\end{table}

The organization of the paper is as follows. In section \ref{GD}, we study the first objective we mentioned above and clarify the geometric destabilization mechanism of gauge fields. In section \ref{Aad}, we study the second objective and reveal the geometric structure of general attractors after the destabilization. In section \ref{examples}, we provide several examples.
In section \ref{perturbation}, we study the primordial fluctuations around these attractors. In section \ref{PPS}, we calculate the primordial power spectrum of the models. The final section is devoted to the conclusion.

\section{Geometric destabilization}\label{GD}
In the case of  multi-scalar fields, the destabilization comes from the negative contribution in the mass term of entropic perturbations. Similarly, a non-trivial geometry of gauge field space has an effect on the conventional slow roll inflation. For simplicity, let us first consider a model with one scalar field and a triplet of one-form gauge fields, where the metric of field space is given by
\begin{equation}\label{f_metric}
G_{ab}(\phi)=
\left(\begin{array}{cc}
    1 &  \\
      & f_{ab}(\phi)
\end{array}\right),
\end{equation}
$a,b=1,2,3$. In general, gauge fields break the rotational symmetry of the theory. However, a specific choice of three $U(1)$ gauge fields allows us to take an isotropic background. In fact, one can identify the internal global $O(3)$ rotation 
of three $U(1)$ gauge fields with the rotation in real three dimensional space 
by choosing the background and the coupling functions as
\begin{equation}\label{bgv2}
A^a_{\ 0}=0,\ \ \ \ \ A^a_{\ i}=\mathbb{A}\delta_{ai}, \ \ \ \ \ f_{ab}=f^2\delta_{ab} \ .
\end{equation}
Apparently, this configuration can be compatible with the isotropic cosmological model. We shall see in the appendix \ref{quadratic_action}, the "mass" term of gauge-field perturbations is independent of the number of gauge fields if $f_{ab}=f^2\delta_{ab}$. Hence the discussion about the instability in this section is also applicable to anisotropic inflation with only one gauge field \cite{Watanabe:2009ct}. In the isotropic configuration of gauge fields, we can simply consider an isotropic flat FLRW cosmological background, i.e., 
\begin{equation}\label{bgv0}
    ds^2=-dt^2+ a^2 (t) \delta_{ij} dx^i dx^j ,
\end{equation}
where $t$ is the cosmic time and $a(t)$ represents a scale factor.  The equations of motion for the homogeneous $\phi$ and $\mathbb{A}$ are
\begin{align}
    &\ddot{\phi}+3H\dot{\phi}+V_{\phi}-3f_{\phi}f\dot{\mathbb{A}}^2a^{-2}=0,\label{eqphi}\\
    &\ddot{\mathbb{A}}+\left(2\frac{\dot{f}}{f}+H\right)\dot{\mathbb{A}}=0, \label{eq}
\end{align}
where the dot denotes differentiation with respect to the cosmological time $t$. The last term of equation of motion of $\phi$ is an additional contribution coming from the non-trivial field space. There is also a geometric term in the equation of motion of gauge fields. If we switch to the conformal time $\tau$, we have $\mathbb{A}''+2(f'/f)\mathbb{A}'=0$. For a trivial geometry $f=$ constant, the gauge field conformally coupled to the FRW background. Hence, $\mathbb{A}$ is an harmonic oscillator and its fluctuations are not excited. While, a non-trivial geometry  provides an effective mass term to the fluctuations.

Let us examine fluctuations of the gauge fields. Thanks to the isotropic configuration of gauge fields, one can
decompose the perturbations of spatial sector of gauge fields as
\begin{equation}
    A^a_{\ i}=a(t)\left(\tilde{Q}+\delta\tilde{Q}\right)\delta_{ai}+a(t)\epsilon_{iab} \partial_b\tilde{U},
\end{equation}
where we have fixed the gauge redundancy of the gauge fields (see appendix \ref{quadratic_action} for details). $\delta \tilde{Q}$ and $\partial\tilde{U}$ are the fluctuations of electric fields and magnetic fields respectively. Here we consider quantities $\mathbb{A}/a(t)$ because we want to discuss the similar mechanism of scalar-field destabilization, and $\tilde{Q}$ transforms exactly like a scalar field under rotation. Moreover, to obtain the similar equations of motion to the scalar field, we introduce new variables of perturbations
\begin{equation}
    \delta Q\equiv\sqrt{2}f\delta \tilde{Q},\ \ \ \ \ \ U_i\equiv\sqrt{2}f \partial_i\tilde{U}. \label{refefine}
\end{equation}
Then one can obtain the equations of motion of these perturbations by varying the quadratic action after perturbing the background (see appendix \ref{quadratic_action} for details). The equations of motion of gauge-field degrees of freedom $\delta Q$ and $U$ are given by
\begin{align}
    &\delta \ddot{Q}+3H\delta \dot{Q}+\left(M_{QQ}+\frac{k^2}{a^2}\right)\delta Q+M_{Q\phi}\delta\phi+\bar{M}_{Q\phi}\delta\dot{\phi}=0,\label{Qeom}\\
    &\ddot{ U}_i+3H\dot{ U}_i+\left(M_{UU}+\frac{k^2}{a^2}\right) U_i=0,
\end{align}
where 
\begin{align}
    M_{QQ}&\equiv 2\epsilon_E\left(3-\epsilon_H\right)H^2-\left(\frac{\ddot{f}}{f}+H\frac{\dot{f}}{f}-2H^2+\epsilon_HH^2\right),\label{MQ2}\\
    M_{UU}&\equiv -\left(\frac{\ddot{f}}{f}+H\frac{\dot{f}}{f}-2H^2+\epsilon_HH^2\right)
    .\label{MU2}
\end{align}
Here, we defined the slow roll parameter $\epsilon_{H}\equiv-\dot{H}/H^2$ and $\epsilon_E\equiv f^2\dot{\mathbb{A}}^2/(M_{\text{pl}}aH)^2$.
The coefficients $M_{Q\phi}$ and $\bar{M}_{Q\phi}$ describe interactions
between scalar field and electric field, which is related to the energy density of gauge fields $\epsilon_E$ as is shown in appendix \ref{quadratic_action}.
 If we assume that the energy of gauge fields is much smaller compared to that of the inflation, i.e., $\epsilon_E\ll 1$, the first term in (\ref{MQ2}) can be ignored
 and we have $M_{QQ}\simeq M_{UU}$.

Similarly, we can also consider the case of the two-form gauge fields. We also consider a triplet  of two-form gauge fields with field space
\begin{equation}\label{h_metric}
G_{ab}(\phi)=
\left(\begin{array}{cc}
    1 &  \\
      & g_{ab}(\phi)
\end{array}\right),
\end{equation}
$a,b=1,2,3$. For any two-form gauge field $A^a_{\ \mu\nu}$ one can spatially dualize the spatial parts of it as $A^a_{\ ij}\equiv\epsilon_{ijk}B^a_{\ k}$. Hence we can diagonalize $B^a_{\ k}$ exactly like the one-form gauge fields \cite{Bento:1992wy,Germani:2009iq}. Then the ansatz of the two-form gauge fields is given by
\begin{equation}\label{bg2v2}
    A^a_{\ 0i}=0,\ \ \ \ \ \ B^a_{\ k}=\mathbb{B}\delta_{ai},\ \ \ \ \ \ g_{ab}=g^2\delta_{ab}.
\end{equation}
Using the above configuration and the metric (\ref{bgv0}), we can obtain the background equations of motion for the homogeneous $\phi$ and $\mathbb{B}$ as
\begin{align}
    &\ddot{\phi}+3H\dot{\phi}+V_{\phi}-3g_{\phi}g \dot{\mathbb{B}}^2a^{-4}=0,\\
    &\ddot{\mathbb{B}}+\left(2\frac{\dot{g}}{g}-H\right)\dot{\mathbb{B}}=0.
\end{align}
The last term in equation of $\phi$ and the second term in equation of $\mathbb{B}$ come from the geometry of field space. Again, we use the helicity decomposition of the perturbations of gauge fields and fix their gauge (see appendix \ref{2quadratic_action}). Then we have the spatial sector of perturbations
\begin{equation}
    B^a_{\ k}=a(t)\left(\tilde{P}+\delta\tilde{P}\right)\delta_{ak},
\end{equation}
where $\tilde{P}$ is the background field. In the two-form case we have only one dynamical scalar perturbation $\delta\tilde{P}$ of the gauge fields. Similarly, to obtain the equations of motions of $\delta\tilde{P}$ that contains the effective mass squared we need, one can define the new variable
\begin{equation}
    \delta P\equiv \frac{g}{a}\delta\tilde{P}.
\end{equation}
Taking the variation of the quadratic action (see Appendix \ref{2quadratic_action}), the equations of motion of two-form gauge fields perturbations $\delta P$ is given by
\begin{equation}
    \delta \ddot{P}+3H\delta\dot{ P}+\left(M_{PP}+\frac{k^2}{a^2}\right)\delta P+M_{P\phi}\delta\phi+\bar{M}_{P\phi}\delta\dot{\phi}=0,\label{Peom}
\end{equation}
where
\begin{equation}\label{MP2}
    M_{PP}\equiv 2\epsilon_B\left(2-\epsilon_H\right)H^2-\left(\frac{\ddot{g}}{g}-\frac{H\dot{g}}{g}-2H^2+\epsilon_HH^2\right)
\end{equation}
and $\epsilon_B\equiv g^2\dot{\mathbb{B}}^2/(M_{\text{pl}}^2a^4H^2)$. 
The coefficients $M_{P\phi}$ and $\bar{M}_{P\phi}$ are related to the energy density of two-form gauge fields $\epsilon_B$.

\begin{figure}[tbp]
\centering
\includegraphics[scale=0.61]{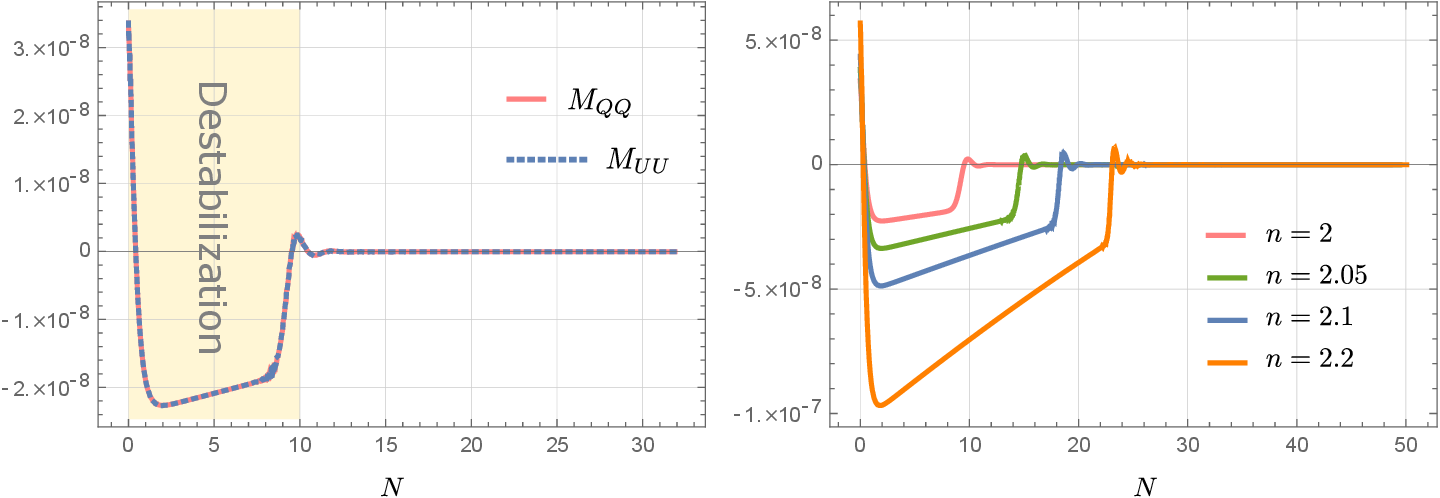}
\caption{\label{fig:MU}($Left$) The effective mass squared of one-form gauge-field perturbations for $f(\phi)=\exp{(\phi^2)}$. $M_{QQ}$ and $M_{UU}$ coincide during inflation and are negative in the unstable inflationary phase (yellow region). ($Right$) The effective mass squared $M_{UU}$ is depicted for $f(\phi)=\exp{(\phi^n)}$ with various  $n=\{2,2.05,2.1,2.2\}$. The mass squared converging to zero implies the stabilization. We chose the potentials $V(\phi)=m^2\phi^2/2$ with $m=10^{-5}$ and set $M_{\text{pl}}=1$.}
\end{figure}
To see how geometric effects of field space destabilize inflation, we assume at the beginning of the evolution, the energy density of the gauge fields is negligible, i.e., $\epsilon_E=\epsilon_B=0$. During this period the interaction between gauge-field modes and scalar-field modes can be ignored and hence have $M_{Q\phi}=\bar{M}_{Q\phi}=M_{P\phi}=\bar{M}_{P\phi}=0$. In this period $M_{QQ}$ and $M_{PP}$ can be regarded as the mass squared of modes $\delta Q$ and $\delta P$ respectively on super-horizon scale $k\ll aH$. For non-trivial geometry of gauge-field space $f$ and $g$, the gauge-field perturbations may become tachyonic, i.e., 
\begin{equation}\label{m2}
    M^2_{A}\equiv M_{QQ} \big{|}_{\epsilon_E=0}<0,\ \ \ \ \ \ M^2_{B}\equiv M_{PP}\big{|}_{\epsilon_B=0}<0,
\end{equation}
hence destabilizes the background evolution. If the evolution of metric $f$ is fast enough ($\dot{f}/f\gg H$), the instability can take place well before the end of inflation. We have showed some examples in Figure \ref{fig:MU} which are unstable at the beginning of inflation. Notably, this condition of instability do not rely on the number of scalar fields because the mass term is only  determined by the geometry of gauge-field space and initial conditions of the destabilization. We shall provide some one-scalar-field and two-scalar-field examples in section \ref{examples}.

\section{Geometric structure of multi-form-field space}\label{Aad}
The geometry in the space of gauge fields
makes the conventional slow-roll inflationary solutions unstable. As in the case of two-scalar-field inflation
such as sidetracked attractors \cite{Garcia-Saenz:2018ifx} or hyperbolic attractors \cite{Brown:2017osf}, this does not mean the inflation end, rather it indicates a new stable attractor~\cite{Watanabe:2009ct}.
Let us look at the attractor when $\ddot{\phi}\ll H\dot{\phi}$, $\epsilon_H\ll 1$ and $\epsilon_E\ll 1$ for one-form gauge fields and $\epsilon_B\ll1$ for two-form gauge fields. In this section, we set $M_{\text{pl}}=1$.
\begin{figure}[tbp]
\centering
\includegraphics[scale=0.73]{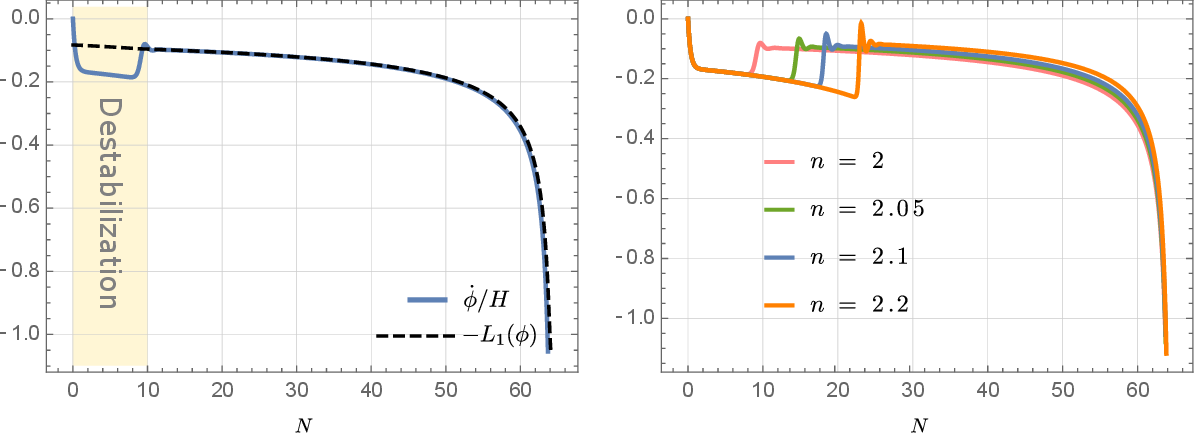}
\caption{\label{fig:Attractors}($Left$) The evolution of $\dot{\phi}/H$ (blue solid curve) for one-form case is depicted. After the destabilization  occurs (yellow region), it will move on to a new attractor given by $-L_1(\phi)$(black dashed curve). ($Right$) The evolution of $\dot{\phi}/H$ for $f(\phi)=\exp{(\phi^n)}$ with $n=\{2,2.05,2.1,2.2\}$ are plotted. We chose the potentials $V(\phi)=m^2\phi^2/2$ with $m=10^{-5}$ and set $M_{\text{pl}}=1$.}
\end{figure}
\subsection{One-form gauge fields}
The attractors of the second phase should have non-decay fluctuations of gauge fields in the power spectrum. Let us write down the equation of motion (\ref{Qeom}) with conformal time $d\tau=dt/a$
\begin{equation}
    \delta{Q}''+2aH\delta{Q}'+\left(2a^2H^2-\frac{f''}{f}\right)\delta{Q}=0\label{Qeomp},
\end{equation}
where we have ignored slow roll parameters. To generate non-decay power spectrum of electric fields, the new attractors should have $f\sim \tau^2$ so that the equations of fluctuations $a\delta Q$ and $aU$ are the same as that of massless scalar fields in de Sitter space \cite{Bartolo:2012sd}. During inflation where $1/\tau\simeq -aH$, from (\ref{Qeomp}) we find the attractor after end of the destabilization always has\footnote{Although we here only show the cases that $f=\exp{(\phi^n)}$ where $n\simeq2$. We have also tested the cases $f\sim \exp{(a\phi^4)}, \exp{(b\phi^3)}, \phi^c$, all of their $M_{QQ}$ transition to almost zero after destabilization.}
\begin{equation}
    \ \ \ \ M_{QQ}\simeq M_{UU}\simeq 0\ \ \ \ \ \ \text{(After end of destabilization)}
\end{equation}
(see Figure \ref{fig:MU}). From the definitions (\ref{MQ2}) and (\ref{MU2}) we have
\begin{equation}\label{masslesseq}
     \frac{M_{QQ}}{H^2}\simeq\frac{M_{UU}}{H^2}\simeq-\frac{f_{\phi\phi}}{f}\left(\frac{\dot{\phi}}{H}\right)^2-\frac{f_{\phi}}{f}\left(\frac{\dot{\phi}}{H}\right)+2\simeq 0,
\end{equation}
where we have neglected the slow roll terms $\ddot{\phi}$, $\epsilon_H$ and $\epsilon_E$.   It is straightforward to solve this equation with respect to $\dot{\phi}/H$ as
\begin{equation}\label{g_attractors}
    \dot{\phi}=-HL_{1\pm}(\phi),
\end{equation}
where we have defined
\begin{equation}\label{L}
    L_{1\pm}(\phi)\equiv -\frac{-f_{\phi}/f\pm \sqrt{\left(f_{\phi}/f\right)^2+8f_{\phi\phi}/f}}{2f_{\phi\phi}/f}.
\end{equation}

In Figure \ref{fig:Attractors} ($Left$), we numerically solved the time evolution and depicted  $\dot{\phi}/H$ and $L_{1-}(\phi)$. The time evolution of scalar field $\dot{\phi}/H$ and the geometrical scale $L_{1-}(\phi)$ given by (\ref{L}) shows good agreement in the attractor region. We note that these attractors are determined by the metric of gauge-field space $f_{ab}(\phi)$. In other words,  the shape of potential is irrelevant to the time evolution of $\dot{\phi}/H$
in the attractor region. We plotted several trajectories for various $f(\phi)$ in Figure \ref{fig:Attractors} ($Right$). 
\begin{figure}[tbp]
\centering
\includegraphics[scale=0.59]{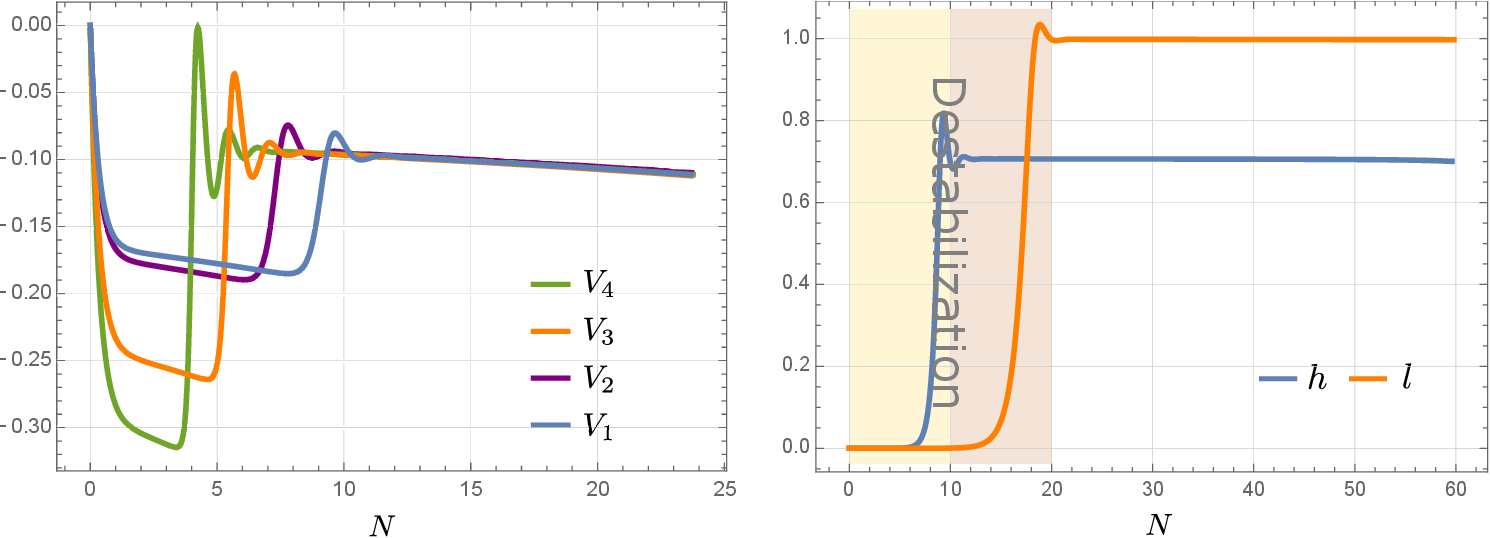}
\caption{\label{fig:3}($Left$) We plotted the evolution of $\dot{\phi}/H$ for different potentials $V_i=\{m_1^2\phi^2/2,m_2^2(\phi^2-\phi)/2,m_3^2(\phi^3+\phi^2)/2,m_4^2(\phi^4+10\phi^3)/2\}$, where $m_i=\{10^{-5},10^{-6},10^{-5},8\times 10^{-7}\}, (i=1,2,3,4)$. We used $f=\exp{(\phi^2)}$ and set $M_{\text{pl}}=1$. 
They converge to the same attractor after the transition.   ($Right$) The evolution of $h\equiv\sqrt{\epsilon_E}/L_1$ and $l\equiv\sqrt{\epsilon_B}/L_2$ are plotted for $f=g=\exp{(\phi^2)}$ and $V=m^2\phi^2/2$, with $m=10^{-5}$. 
In this case, $h$ and $l$ are nearly constants. }
\end{figure}

On the other hand, the shape of potential plays an important  role in determining the conditions for destabilization. Before the destabilization occurs, the system was in the conventional slow roll inflation: $3H\dot{\phi}\simeq -V_{\phi}$. Substituting it into (\ref{m2}), we immediately see the slow roll inflation become unstable when
\begin{equation}\label{condition1}
    \frac{V_{\phi}}{3H}>HL_{1-}(\phi)\ \ \ \ \text{or}\ \ \ \ \frac{V_{\phi}}{3H}<HL_{1+}(\phi).
\end{equation}
After that, the system shows a transition to the attractor (\ref{g_attractors}). In Figure \ref{fig:3} ($Left$), we show the attractor does not depend on the shape of potentials. Note that in the oroginal paper of anisotropic inflation \cite{Watanabe:2009ct}, the metric in field space is chosen as $f=\exp{(2c\int V/V_{\phi}d\phi})$ which depends on the potential. As is emphasized in \cite{Soda:2012zm}, however,
 this choice is not mandatory. Here we clearly shows the choice of $f$ needs not depend on the potential. The destabilization occurs as long as (\ref{condition1}) is satisfied.

\subsection{Two-form gauge fields}
\begin{figure}[tbp]
\centering
\includegraphics[scale=0.57]{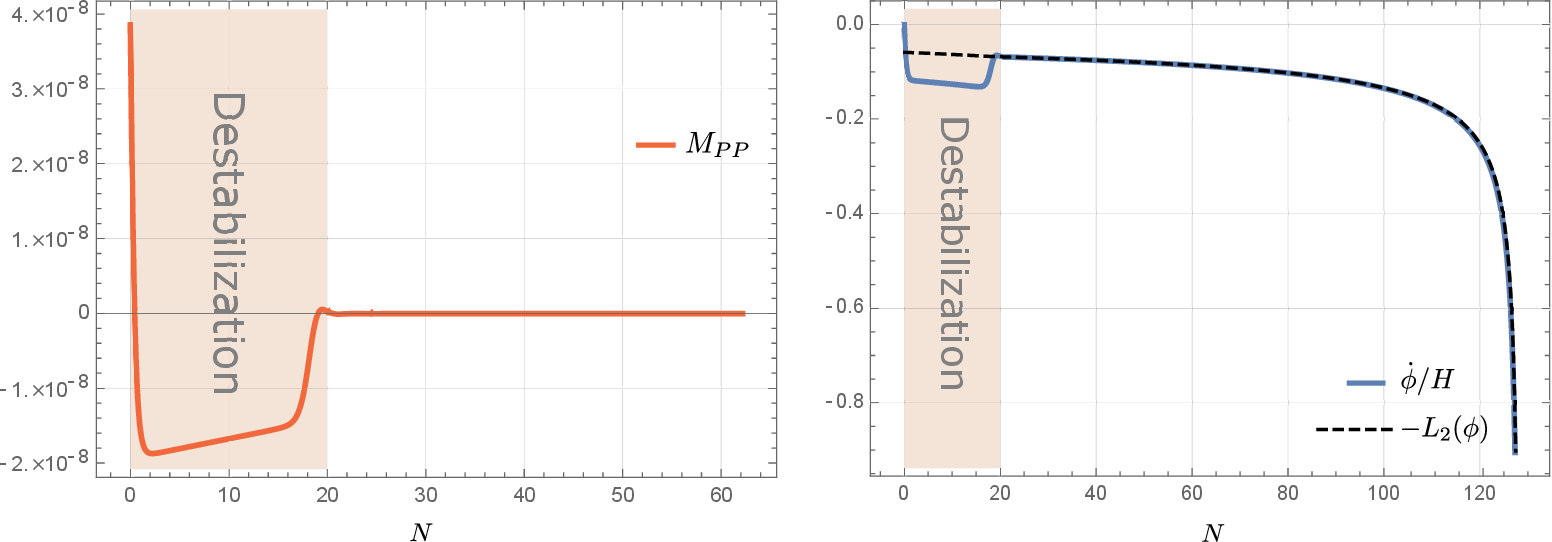}
\caption{\label{fig:4}($Left$) The effective mass squared of two-form gauge-field perturbations for $h(\phi)=\exp{(\phi^2)}$. $M_{PP}$ is negative in the unstable inflationary phase (red region). ($Right$) The evolution of $\dot{\phi}/H$ (blue solid curve) for two-form case. 
Due to the destabilization (red region), it shows a transition to a new attractor given by $-L_2(\phi)$(black dashed curve). We set $M_{\text{pl}}=1$.}
\end{figure}

For the two-form gauge fields, to produce the non-decay power spectrum of the gauge fields, we can see in the equation of motion (\ref{Peom}) with conformal time
\begin{equation}
    \delta P''+2aH\delta P'+\left(2a^2H^2-\frac{g''}{g}+2aH\frac{g'}{g}\right)\delta P=0,\label{Peomp}
\end{equation}
where we have ignored slow roll parameters. If we are in the de Sitter space during inflationary period where $1/\tau\simeq -aH$, the equation of fluctuations $a\delta P$ is the same as that of massless scalar fields if the field metric $g\sim \tau$ \cite{Ohashi:2013qba}. From (\ref{Peomp}), we find the perturbation of two-form gauge fields in the attractor region  has 
\begin{equation}
     \ \\ \ \ M_{PP}\simeq 0\    ,
\end{equation}
which can be verified numerically as shown in Figure \ref{fig:4}. Form the definition of $M_{PP}$ we have
\begin{equation}\label{masslesseq2}
    \frac{M_{PP}}{H^2}\simeq-\frac{g_{\phi\phi}}{g}\left(\frac{\dot{\phi}}{H}\right)^2+\frac{g_{\phi}}{g}\left(\frac{\dot{\phi}}{H}\right)+2\simeq 0 ,
\end{equation}
where we again ignored the slow roll parameters. We note the second term has the opposite sign to that of (\ref{masslesseq}). We can solve this equation and yield two solutions
\begin{equation}\label{2g_attractors}
    \dot{\phi}= -HL_{2\pm},
\end{equation}
where $L_{2\pm}$ are defined by
\begin{equation}\label{L2}
    L_{2\pm}(\phi)\equiv -\frac{g_{\phi}/g\pm \sqrt{\left(g_{\phi}/g\right)^2+8g_{\phi\phi}/g}}{2g_{\phi\phi}/g}.
\end{equation}
From Figure \ref{fig:4}, we see the evolution of $\dot{\phi}/H$ and the geometrical scale $L_{2-}$ shows good agreement.
Note that  the sign of $g_{\phi}/g$ in (\ref{L2}) is opposite to that in (\ref{L}). We also have two solutions for a given $g(\phi)$.

Now, we  can deduce the conditions for destabilization 
of the conventional slow roll inflationary phase. Same as the one-form gauge fields, after substituting slow roll solution $\dot{\phi}/H\simeq-V_{\phi}/(3H^2)$ into (\ref{m2}) we obtatin
\begin{equation}\label{condition2}
    \frac{V_{\phi}}{3H}>HL_{2-}(\phi)\ \ \ \ \text{or}\ \ \ \ \frac{V_{\phi}}{3H}<HL_{2+}(\phi).
\end{equation}
Similar to the hyperbilic inflation, when the potential is steep engouh, the conventional slow roll solution becomes unstable.

We have found that there are two attractors for a given $f(\phi)$ (or $g(\phi)$). However, not both of them are well-behaved solutions for inflation. For some of these solutions, the coupling of scalar fields and gauge fields is strong at the early time of inflation. Notice that $f(\phi)$ is inversely proportional to the effective coupling constant \cite{Martin:2007ue,Demozzi:2009fu,Kanno:2009ei} and we typically have  $f(\phi_e)\sim O(1)$ at the end of inflation. If $f(\phi)$ grows with the scale factor, the effective coupling is incredibly large at the beginning of inflation. Hence we are in the strong coupled regime where the perturbation method is not reliable. If we assume  the system is in the conventional slow roll inflation before the destabilization, we should avoid strong coupling with gauge fields. We classifiy the strongly coupling and weakly coupling attractors for differnt choice of metric $f(\phi)$ in the Appendix \ref{class}. In this paper, we only consider the right-to-left rolling ($L_>0$) hence in the rest of discussion we only consider one of these attractors, namely, the weakly coupled $L_{-}(\phi)>0$ case.
Hereafter, the minus subscript will be omitted.  

\subsection{Geometry of attractors}\label{Goa}
Let us discuss slow roll variables in the attractor phase for both one-from case and two-form case. First, in these attractors, the energy density of one-form gauge fields and two-form gauge fields are given by 
\begin{align}
\rho_A\equiv \frac{3f^2\dot{\mathbb{A}}^2}{2a^2}=\frac{3}{4}\left(2\epsilon_H-L_1^2\right)H^2, \label{E-energy}\ \ \ \ \ 
\rho_B\equiv \frac{3g^2\dot{\mathbb{B}}^2}{2a^4}=\frac{3}{2}\left(2\epsilon_H-L_2^2\right)H^2 \ .
\end{align}
These quantities are also slowly varying during the inflation because $\epsilon_H\equiv-\dot{H}/H^2$, $H$ and $L$ are all slowly varying,
\begin{equation}
|\eta_L|\equiv\bigg|\frac{\dot{L}}{HL}\bigg|=\bigg|\frac{\dot{\sqrt{\epsilon}}}{H\sqrt{\epsilon}}\bigg|=|\eta+\epsilon_H|\ll 1,\label{lchang}
\end{equation}
where $\sqrt{\epsilon}\equiv\dot{\phi}/(\sqrt{2}H)$\footnote{The root is just a symbol and does not mean that $\sqrt{\epsilon}$ is positive. In this paper we always consider the right-to-left rolling $\phi$ so $\sqrt{\epsilon}<0$}. 
The Hubble constant given by $3H^2\simeq V$ is slowly varying. So this attractor can provide enough e-folding number $N(\phi)=-\int^{\phi}{d\phi/L(\phi)}$ to solve the horizon problem. Differentiating $3H^2\simeq V$ with respect to time twice and using (\ref{g_attractors}),  we obtain other slowly varying parameters 
\begin{align}
    \epsilon_V&\equiv\frac{L(\phi)V_{\phi}}{V}=2\epsilon_H\ll 1, \label{epsilonL}\\
    |\eta_V|&\equiv\bigg|\frac{L(\phi)V_{\phi\phi}}{V_{\phi}}\bigg|=|3\epsilon_H+\eta-\eta_H|\ll 1,\label{etaL}
\end{align}
where $\eta\equiv\ddot{\phi}/(H\dot{\phi})$ and $\eta_H\equiv \dot{\epsilon}_H/(H\epsilon_H)$. Moreover, we define a parameter of the ratio of energy density of gauge fields and kinetic energy of scalar field, i.e.,
\begin{align}
    &h\equiv\frac{\sqrt{\epsilon_E}}{L_1(\phi)}=\sqrt{\frac{1}{2}\left(\frac{V_{\phi}}{3L_1H^2}-1\right)}=\sqrt{\frac{1}{2}\left(\frac{\epsilon_V}{L_1^2}-1\right)}\label{h},\\
    &l\equiv\frac{\sqrt{\epsilon_B}}{L_2(\phi)}=\sqrt{\frac{V_{\phi}}{3L_2H^2}-1}=\sqrt{\frac{\epsilon_V}{L_2^2}-1}\label{l},
\end{align}
where we have used (\ref{E-energy}) and Friedmann equation in the second equality. Here $\epsilon_E\equiv f^2\dot{\mathbb{A}}^2/(M_{\text{pl}}aH)^2$ and $\epsilon_B\equiv g^2\dot{\mathbb{B}}^2/(M_{\text{pl}}^2a^4H^2)$. From the instability conditions (\ref{condition1}), we see the slow roll inflation is unstable (stable) when $h$ and $l$ are real (imaginary) because of the exponential enhancement of gauge fields. We also calculate the 
growth rate of $h$. Using $\epsilon_V=2\epsilon_H$, we have
\begin{align}\label{hchange}
    \eta_h\equiv\frac{\dot{h}}{Hh}=\frac{1+2h^2}{4h^2}\left(\eta_H-2\eta_L\right),\ \ \ \ \ \ 
    \eta_l\equiv\frac{\dot{l}}{Hl}=\frac{1+l^2}{2l^2}\left(\eta_H-2\eta_L\right)
\end{align}
We find if $h$ and $l$ are not too small, i.e., $h>\mathcal{O}(1)$ and $l>\mathcal{O}(1)$, $\eta_h$ and $\eta_l$ are in the same order of the slow roll parameters  $\eta_H-2\eta_L$.  Hence, $h$ and $l$ are  slowly varying. On the other hand, if $h\ll 1$ and $l\ll 1$, the coefficient of (\ref{hchange}) $\sim 1/(h^2)$ becomes very large. However, in these cases we have $\epsilon_H\simeq \dot{\phi}^2/(2H^2)=L^2/2$ so $\eta_H\simeq 2\eta_L$. Hence $\eta_h$ and $\eta_l$ can still be small. In Figure \ref{fig:3}, we show an example of slowly varying parameter $h$. It can be regarded as a constant at the leading order of slow roll approximation\footnote{In our discussion the leading order of slow roll approximation means all slow roll parameters $\epsilon$ vanish but $\sqrt{\epsilon}$ do not. It is also valid for the calculations of perturbations below.}, which will be important in calculating the perturbations, see Section \ref{deltaphi}.

On the other hand, from the equation of motion of $\phi$ (\ref{eqphi}), (\ref{epsilonL}) and Friedmann equation, the energy density $\rho_A$ and $\rho_B$ can be represented as $\rho_A=3f\left(2\epsilon_H-L_1^2\right)H^2/\left(2f_{\phi}L_1\right)$ and $\rho_B=3g\left(2\epsilon_H-L_2^2\right)H^2/\left(2g_{\phi}L_2\right)$, respectively. Comparing these with (\ref{E-energy}), we obtain one of our main results
\begin{equation}
    \frac{f_{\phi}}{f}=\frac{2}{L_1(\phi)}, \ \ \ \ \ \ \ \ \ \  \frac{g_{\phi}}{g}=\frac{1}{L_2(\phi)}. \label{geometry}
\end{equation}
These are differential equations of $f(\phi)$ and $g(\phi)$ respectively. Solving these equations yields the geometry of field space for these attractors. However, for inflation where $\dot{\phi}/H\lesssim1$, $L_{\pm}(\phi)$ should be sub-Planckian $L(\phi)\lesssim 1$ and slowly varying during inflation (see (\ref{lchang})). In other words, 
if we consider the leading order of the slow roll approximation on a de Sitter background where $\epsilon_H\simeq\eta\simeq 0$, $L$ is almost a constant during inflation. Then the metric of field space (\ref{f_metric}) becomes
\begin{equation}\label{hyperbolic}
    f(\phi)\sim e^{2\phi/L_1}, \ \ \ \ \ \ \ \ g(\phi)\sim e^{\phi/L_2},
\end{equation}
which are hyperbolic type with sub-Planckian radius $L_1/2$ and $L_2$, respectively. In other worlds, no matter what we choose $f(\phi)$ and $g(\phi)$ as a set up, as long as there occurs a transition to this attractor, the field space becomes
a hyperbolic space at the leading order of the  slow roll approximation. 
Thus, we have revealed the geometric structure of the attractors.
This is why the attractor looks similarly to one of the hyperbolic inflation. Later, we will use the method which was useful to investigate hyperbolic inflation 
to calculate the perturbations of this model in section \ref{perturbation}.

We give some comments on this one-form model compared to the hyperbolic inflation. The two-form model is similar. First, in this model we did not expect that the geometry of field space we start is a hyperbolic space. We considered general metric $f$ for which the geometrical destabilization occurs. Second, there should be a mass scale $M$ in the geometry of field space $f$ to characterize the destabilization. A scale $L_h$ of the irrelevant operator $\sim L_h^2\sinh^2(\phi/L_h)(\nabla\theta)^2$ in hyperbolic inflation is expected to be lower than the UV cutoff of the effective theory. It characterizes the destabilization because it is related to the curvature of field space. Moreover, in contrast to hyperbolic inflation, $L$ is running with rolling of the scalar field. But it does not change significantly before the end of inflation, thanks to the slow rolling. Finally, in the hyperbolic inflation, to make scalar slowly rolling, the potential force $V_{\phi}$ should be balanced with the centrifugal force $L_h\sinh^2{(\phi/L_h)}\dot{\theta}^2$. This means the energy of angular motion should be much larger than the kinetic energy of scalar field so that the ratio satisfies the inequality $h_h\equiv\sqrt{V_{\phi}/(L_hH^2)-9}\gg 1$. This ratio is similar to the ratio $h$ in our attractor defined by (\ref{h}), i.e., the ratio of the energy of gauge fields and the kinetic energy of the scalar field. However, we do not need to satisfy the inequality $h\gg 1$. A known example is the metric $f(\phi)=\exp{\left(2c\int V/V_{\phi}d\phi\right)}$. 
In this case, if we consider the isotropic configuration of gauge fields , we still have $-V_{\phi}/(3H\dot{\phi})\simeq V_{\phi}/(3H^2L)=c$. If we choose $c\sim\mathcal{O}(1)$ we have $h\sim\mathcal{O}(1)$. Nevertheless, the ratio $h$ is not arbitrary and we shall 
discuss the constraint from observations in section \ref{perturbation}.

\section{Examples}\label{examples}
In this section, we show some specific examples for the geometric destabilization. We both consider the single-field and two-field models coupled with gauge fields. In the two-field cases, we assume the shift symmetry of the second field so that the metric of field space only contains inflaton $\phi$. In particular, our examples demonstrate that the destabilization does not depend on the number of fields. 
\subsection{Power-law inflation}
We consider the case where the slow roll parameters are constants during inflation. 
We first take a single-field model with an exponential type of gauge-field metric and potential
\begin{equation}
G_{ab}(\phi)=
\left(\begin{array}{cc}
    1 &  \\
      & e^{\rho\phi}\delta_{ab}
\end{array}\right),\ \ \ \ \ \ \ \ V(\phi)=e^{\lambda\phi},
\end{equation}
where $\lambda$ and $\rho$ are positive parameters. In this case, the attractor is a fixed point in the parameter space of $\lambda$ and $\rho$. Hence inflation never ends. When gauge fields are negligible for the  background, we have an exact power-law attractor for this system \cite{Halliwell:1986ja,Copeland:1997et}
\begin{equation}
    \alpha=\zeta\log t,\ \ \ \ \ \ \ \ \phi=\xi\log t+\phi_0,
\end{equation}
where $\phi_0$ is an initial value of the scalar field and we defined
\begin{equation}
    \zeta=\frac{2}{\lambda^2},\ \ \ \ \ \ \ \ \xi=-\frac{2}{\lambda}.\label{onepl}
\end{equation}
However, this attractor is unstable if the perturbation modes of gauge field $\delta Q$ and $U$ become tachyonic during the inflationay period. In other words, when
\begin{align}
   \frac{M^2_{A}}{H^2}\bigg{|}_{\epsilon_E=0}&=-\frac{1}{\zeta^2}\left(\rho^2\xi^2-\rho\xi+\zeta\rho\xi-2\zeta^2+\zeta\right)\nonumber\\
   &=-\frac{1}{2}\left(\rho\lambda+1\right)\left(\lambda^2+2\rho\lambda-4\right)<0,
\end{align}
the attractor is unstable. Notably here we used (\ref{MQ2}) where the slow-roll parameters are not ignored. In the second equality, we have used the solutions (\ref{onepl}). Hence, for positive $\lambda$ and $\rho$, the condition for the destabilization is given by $\lambda^2+2\rho\lambda-4>0$, which is consistent with \cite{Chen:2021nkf}. The system will evolve to a new anisotropic power-law attractor which is stable \cite{Kanno:2010nr}. As we mentioned before, the destabilization condition does not depend on the number of fields. 

We can also discuss the multi-scalar-field case. Here we consider the hyperbolic inflationary model with two scalar fields ($\phi,\theta$) and isotropic gauge fields background, where the metric of field space is given by 
\begin{equation}
    G_{ab}(\phi)=
\left(\begin{array}{ccc}
    1 & & \\
      & \frac{L^2}{4}e^{2\phi/L} & \\
      & & e^{\rho\phi}\delta_{ab}
\end{array}\right),\ \ \ \ \ \ \ \ V(\phi)=e^{\lambda\phi},
\end{equation}
where $\phi$ is a radial field and $\theta$ is an massless angular field. When gauge fields are negligible for the background, there exists an exact attractor with non-geodesic trajectory (turn rate $\neq 0$) in the scalar-field space \cite{Mizuno:2017idt}
\begin{equation}
    \zeta=\frac{2+L\lambda}{3L\lambda},\ \ \ \ \ \ \ \  \xi=-\frac{2}{\lambda}\label{twopl}.
\end{equation}
The perturbation modes become tachyonic when
\begin{equation}
    \frac{M^2_{A}}{H^2}\bigg{|}_{\epsilon_E=0}=-\frac{6\rho L+L\lambda+2}{\left(2+l\lambda\right)^2}\left(6\rho L+L\lambda-4\right)<0,
\end{equation}
where we have used the attractor (\ref{twopl}) as the initial conditionfor the destabilization. For positive $L$, $\rho$ and $\lambda$, this attractor is unstable because of the backreaction of the gauge fields if $6\rho L+L\lambda-4>0$, which is consistent with \cite{Chen:2021nkf}. In \cite{Chen:2021nkf}, we showed that  the system with one gauge field will evolve to a new stable anisotropic hyperbolic attractor after destabilization.

\subsection{Chaotic inflation}
Now we also consider chaotic inflation with a single-field and an isotropic one-form gauge field, where the metric of field space and potential are given
\begin{equation}
    G_{ab}(\phi)=
\left(\begin{array}{cc}
    1 &  \\
      & e^{c\phi^2/2}
\end{array}\right),\ \ \ \ \ \ \ \ V(\phi)=\frac{1}{2}m^2\phi^2,
\end{equation}
and $c$ is a positive constant. This is the anisotropic inflation where the metric $f$ can be determined by setting $f=\exp{(2c\int V/V_{\phi}d\phi})$ \cite{Watanabe:2009ct}. At the beginning of inflation, the gauge field can be ignored so that the attractor is the conventional slow-roll inflation
\begin{equation}
    3H^2\simeq \frac{1}{2}m^2\phi^2,\ \ \ \ \ \ \ \ 3H\dot{\phi}\simeq -m^2\phi,\label{sla}
\end{equation}
which yields $\dot{\phi}\phi\simeq -2H$. Using the above equations, we obtain 
\begin{equation}
     \frac{M^2_{A}}{H^2}\bigg{|}_{\epsilon_E=0}\simeq-\frac{f_{\phi\phi}}{f}\left(\frac{\dot{\phi}}{H}\right)^2-\frac{f_{\phi}}{f}\left(\frac{\dot{\phi}}{H}\right)+2
     \simeq -2(2c+1)(c-1)<0,
\end{equation}
where we have ignored slow rolling quantities $2\epsilon\equiv \dot{\phi}^2/H^2\ll 1$ and $\ddot{\phi}\ll H\dot{\phi}$. It is easy to see the backreaction of gauge fields occurs if $c>1$. From (\ref{g_attractors}) we have new attractor $\dot{\phi}/H=-L\simeq-2/(c\phi)$. Using $3H^2\simeq V$ we have $3H\dot{\phi}=-m^2\phi/c$, which is consistent with \cite{Watanabe:2009ct}. In other words, if $c>1$ the conventional  slow roll attractor is unstable from the beginning of the evolution and makes a transition to a new attractor when the gauge fields grow sufficiently. Again, the destabilization does not depend on the number of fields. Hence the new inflation phase still occurs for the isotropic background \cite{Firouzjahi:2018wlp,Gorji:2020vnh}.

For one two-form gauge field, where the metric of field space is given by
\begin{equation}
    G_{ab}(\phi)=
\left(\begin{array}{cc}
    1 &  \\
      & e^{c_2\phi^2/4}
\end{array}\right) ,
\end{equation}
we have  anisotropic inflation~\cite{Ohashi:2013mka}. Again, at the beginning, the energy density of two-form gauge field can be ignored, so the attractor is given by (\ref{sla}), which implies $\dot{\phi}\phi\simeq-2H$.  Substituting the slow roll solution into (\ref{masslesseq2}), we immediately have
\begin{equation}
     \frac{M^2_{B}}{H^2}\bigg{|}_{\epsilon_B=0}\simeq-\frac{g_{\phi\phi}}{g}\left(\frac{\dot{\phi}}{H}\right)^2+\frac{g_{\phi}}{g}\left(\frac{\dot{\phi}}{H}\right)+2
     \simeq -(c_2+2)(c_2-1)<0.
\end{equation}
The instability occurs when $c_2>1$, which is consistent with \cite{Ohashi:2013mka}.

\begin{figure}[tbp]
\centering
\includegraphics[scale=0.72]{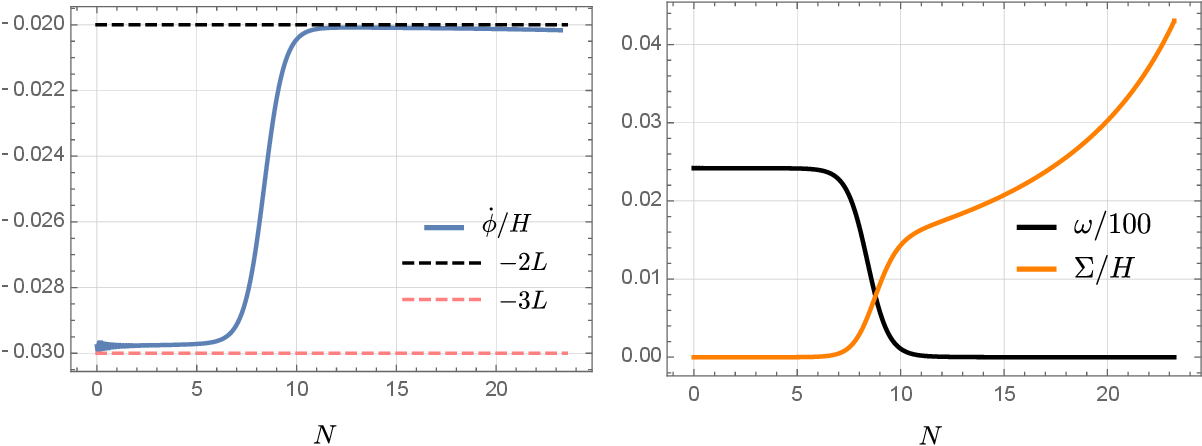}
\caption{\label{fig:2fields}The evolution of $\dot{\phi}/H$ for the hyperbolic field space and the exponential potential is plotted. It shows a transition from $-3L$ to $-2L$ because of the geometrical destabilization, where $L=0.01$. ($Right$) During the transition, the gauge field grows and consequently the anisotropy increases (orange curve) while the turn rate $\omega$ (black curve) of the scalar field drops to zero. }
\end{figure}

We can also consider the case where $\epsilon$ is almost constant during inflation. We here start from a hyperbolic inflation with two scalar fields and a gauge field. 
For the constant $\epsilon$, one can solve differential equations (\ref{masslesseq}) to obtain $f(\phi)\sim e^{\phi/L}$,
hence (\ref{f_metric}) is hyperbolic type with curvature scale $L\equiv-\sqrt{\epsilon}/2$.
The metric in the field space and the potential are given by
\begin{equation}
    G_{ab}(\phi)=
\left(\begin{array}{ccc}
    1 & & \\
      & \frac{L^2}{4}e^{2\phi/L} & \\
      & & \frac{L^2}{4}e^{2\phi/L}
\end{array}\right),\ \ \ \ \ \ \ \ V(\phi)=\frac{1}{2}m^2\phi^2.
\end{equation} 
In contrast to the power-law inflation, the slow roll parameter $\epsilon$ is constant while $\epsilon_H$ is monotonically increasing until $\mathcal{O}(1)$ where inflation ends. 
Similar to hyperbolic inflation, the centrifugal force in field space cancels $V_{\phi}$, i.e., the last two terms in Eq.(\ref{eq}) are balanced to keep $3H\dot{\phi}$ slowly varying: $V_{\phi}=3p_A^2f_{\phi}f^{-3}e^{-4\alpha}$. 
The attractor is given by
\begin{equation}
    \dot{\phi}=-2HL/(1+\eta_L/4)\simeq -2HL,\label{2HL}
\end{equation}
where we have ignored the small parameter $\eta_L/2\equiv LV_{\phi\phi}/V_{\phi}\ll 1$. 

At the beginning of inflation, since the energy of gauge field is negligible, the system is in a hyperbolic inflation attractor $\dot{\phi}\simeq-3HL$. The hyperbolic attractor has non-geodesic trajectory in the scalar-field space. The tangent $T^a$ and normal $N^a$ direction of this trajectory are given by $T^a\equiv\dot{\phi}_{t}^a/|\dot{\phi}_t^a|=(\dot{\phi},\dot{\theta})/|\dot{\phi}_t^a|$ and $N^a
T_a=1$ respectively. One can find the trajectory of hyperbolic attractor has a large turn rate $\omega\equiv\mathcal{D}_tN^a/H$ in the field space, where $\mathcal{D}_t$ is the covariant derivative of the 2-dimensional hyperbolic space.
However this attractor is unstable because
\begin{equation}
    \frac{M^2_{A}}{H^2}\bigg{|}_{\epsilon_E=0}\simeq-\left(2-\frac{3HL}{HL}\right)\left(-1-\frac{3HL}{HL}\right)=-4<0.
\end{equation}
In this regime, the energy density of gauge field $\epsilon_E$, or equivalently the anisotropy $\Sigma/H$ \cite{Watanabe:2009ct} increases. The gauge field has to capture the energy from the angular field $\theta$ to stabilize inflation, which will slow down the turn rate $\omega$ of trajectory in the scalar-field space. Finally, the system will converge to the attractor (\ref{2HL}). Afterwards, the turn rate drops to zero and the anisotropy keeps increasing because $\epsilon_H=\epsilon_E+\epsilon/2$ monotonically increases while $\epsilon$ remains unchanged, see Figure \ref{fig:2fields}.

\section{Primordial fluctuations}\label{perturbation}
In this section, we investigate the dynamics of perturbations around attractors by using the standard connection method (see, for example, \cite{Brown:2017osf,Mizuno:2017idt}).
The details of derivation of the quadratic action can be found in appendix \ref{quadratic_action}. As we mentioned at the end of section \ref{Aad}, unlike the hyperbolic inflation, we do not need to restrict us to a very large $h$ or $l$ because there also exist attractors even when the energy of gauge field is not too large compared to the kinetic energy of scalar field. The situation that $h$ is very small has been studied in some literature before \cite{Yamamoto:2012sq,Funakoshi:2012ym,Firouzjahi:2018wlp,Gorji:2020vnh}, where $h$ is regraded as a perturbation so one can use the standard in-in formalism. Therefore we only consider the dynamics of perturbation in the regime $h,l>\mathcal{O}(1)$ in this section. Because the energy density of gauge fields is not so small, we will use non-perturbative methods to discuss them.

For hyperbolic inflation, there is a new degree of freedom from the second scalar field.
Thus, the power spectrum of curvature perturbations is modified, while the power spectrum of gravitational waves remains unchanged. Hence the modification of tensor-to-scalar ratio comes only from the contribution of curvature perturbations. In this case, the tensor-to scalar ratio will be suppressed exponentially because of the growth of scalar modes inside
the horizon and the quickly angular motion of the second scalar field. However, if we consider the gauge fields, there may be source terms coming from the gauge fields in the equation of gravitational waves. As we mentioned before, the energy of gauge fields can be moderate compared to the kinetic energy of inflaton. Moreover, gauge fields contain vector modes so we should discuss their evolution during inflation. Therefore in this paper we calculate the spectral index of power spectrum of curvature perturbation and tensor-to-scalar ratio for $h,l>\mathcal{O}(1)$. 
Our models should be observationally
distinguishable from the hyperbolic inflation.

\subsection{One-form gauge fields}
Because of the isotropy of gauge fields, we can use helicity decomposition of perturbations
into scalar, vector and tensor parts, see appendix \ref{quadratic_action}. The scalar perturbations are ($A,B,\psi,E,\delta\phi,\mathbb{Y},\delta\mathbb{A},\mathbb{U},\mathbb{M}$), four from metric, one from scalar field and four from gauge fields. The vector perturbations are ($B_i,W_i,\mathbb{Y}_a,\mathbb{U}_a,\mathbb{M}_a$), two from the metric and three from the gauge fields. We choose spatially flat gauge $\psi=E=W_i=0$ and $\mathbb{M}=\mathbb{U}_a=0$ to fix the gauge.  There are only two dynamical tensor perturbations $(w_{ij}, \mathbb{T}_{ij})$. Moreover, ($A,B,\mathbb{Y},B_i,\mathbb{Y}_a$) are non-dynamical so we can solve their equations and plug them into the quadratic action to eliminated them. Finally the only dynamical perturbations we should deal with are $(\delta\phi,\delta\mathbb{A},\mathbb{U},\mathbb{M}_a,w_{ij}, \mathbb{T}_{ij})$. 

\subsubsection{Dynamics of scalar modes}\label{deltaphi}
We first discuss the scalar parts of the perturbations, which contribute to the curvature perturbation on large scale. Because the large enough energy of gauge fields, the curvature perturbation should contain significant amount of modes from the gauge fields. We have redefined the perturbations of gauge fields in (\ref{refefine}): $\delta Q\equiv \sqrt{2}f\delta\mathbb{A}/a$ and $U_i\equiv \sqrt{2}f\partial_i\mathbb{U}/a$. It is convenient to change the time variable to the conformal time $d\tau=dt/a$ and introduce the canonical variables
\begin{equation}\label{nv1}
    \Delta_{\phi}\equiv a\delta{\phi},\ \ \ \ \ \ \Delta_{Q}\equiv a\delta{Q},\ \ \ \ \ \ \Delta_{U}\equiv aU_i
\end{equation}
In the slow roll approximation we can derived the quadratic action of the fluctuations $\Delta_{\phi}$, $\Delta_{Q}$ and $\Delta_{U}$ (see Appendix \ref{2action_sl}). However, $\Delta_{U}$ is an isocurvature mode which does not contribute to the curvature perturbations \cite{Gorji:2020vnh}. Hence we can ignore it from now on. The quadratic action is given by
\begin{align}
    S^{(2)}_{\text{scalar}}=\frac{1}{2}\int d\tau d^3x& \Bigg\{(\Delta_{\phi}')^2-(\partial\Delta_{\phi})^2+(\Delta_{Q}')^2-(\partial\Delta_{Q})^2+\frac{8h^2+2}{\tau^2}(\Delta_{\phi})^2\nonumber\\
    &+\frac{2}{\tau^2}(\Delta_{Q})^2+\frac{16\sqrt{2}h}{\tau^2}\Delta_{\phi}\Delta_{Q}-\frac{8\sqrt{2}h}{\tau}\Delta_{\phi}\Delta_{Q}'
    \Bigg\},\label{action2per}
\end{align}
where $h$ can be obtained from the  background quantities (\ref{h}), which varies slowly during inflation and can be regarded as a constant at leading order in the slow roll approximation. Taking the variation of the action with respect to $\Delta_{\phi}$ and $\Delta_{Q}$ and moving into the Fourier space, we obtain the equations of motion  
\begin{align}
    &\Delta_{\phi}''+\left(k^2-\frac{8h^2+2}{\tau^2}\right)\Delta_{\phi}-\frac{8\sqrt{2}h}{\tau^2}\Delta_{Q}+\frac{4\sqrt{2}h}{\tau}\Delta_{Q}'=0,\label{DeltaPhieq}\\
    &\Delta_{Q}''+\left(k^2-\frac{2}{\tau^2}\right)\Delta_{Q}-\frac{4\sqrt{2}h}{\tau^2}\Delta_{\phi}-\frac{4\sqrt{2}h}{\tau}\Delta_{\phi}'=0.\label{DeltaQeq}
\end{align}
Since $h$ is regarded as a constant, these equations are solvable. 

We are interested in the modes outside the horizon ($|k\tau|\ll 1$), which contribute to the curvature perturbations.
The modes which freeze in or decay on large scales are given by
\begin{align}
    \Delta_{\phi}=&-\frac{3\sqrt{2}}{2h}\frac{c_1}{(-\tau)}+\frac{\sqrt{9-96h^2}-3}{16h}\sqrt{2}c_3(-\tau)^{(1+\sqrt{9-96h^2})/2}\nonumber\label{superphi}\\
    &-\frac{\sqrt{9-96h^2}+3}{16h}\sqrt{2}c_4(-\tau)^{(1-\sqrt{9-96h^2})/2},\\ 
    \Delta_{Q}=&\frac{c_1}{(-\tau)}+c_2(-\tau)^2+c_3(-\tau)^{(1+\sqrt{9-96h^2})/2}+c_4(-\tau)^{(1-\sqrt{9-96h^2})/2}\label{superq},
\end{align}
where $c_1\sim c_4$ are constants of integration. Firstly, for the stable attractor (\ref{g_attractors}), we have $f\sim a^{-2}$ hence $f\dot{\mathbb{A}}/a=$constant. So the gauge field grows as $\mathbb{A}\sim a^3+\Delta_{Q}/(\sqrt{2}f)$. Therefore $c_2$ mode is just a shift of $\mathbb{A}$ and becomes irrelevant rapidly. Secondly, $c_3$ and $c_4$ modes are massive modes and will soon decay away when $h>\sqrt{9/96}\simeq0.3$. Hence the growing modes $c_1$ corresponds to the adiabatic fluctuations, which contribute to the curvature perturbations on large scales. 

The coefficient $c_1$ can be fixed by conditions imposed sub-horizon scales ($|k\tau|\gg 1$). In the sub-horizon regime, we can ignore the $\sim1/\tau^2$ terms in the equations and approximately solve the equations as
\begin{align}
    \Delta_{\phi}=&iC_1 e^{ik\tau+i2\sqrt{2}h\log{|k\tau|}}-iC_2e^{ik\tau-i2\sqrt{2}h\log{|k\tau|}}+iC_3e^{-ik\tau-i2\sqrt{2}h\log{|k\tau|}}\nonumber\\
    &-iC_4e^{-ik\tau-i2\sqrt{2}h\log{|k\tau|}},\\
    \Delta_{Q}=&C_1e^{ik\tau+i2\sqrt{2}h\log{|k\tau|}}+C_2e^{ik\tau-i2\sqrt{2}h\log{|k\tau|}}+C_3e^{-ik\tau+i2\sqrt{2}h\log{|k\tau|}}\nonumber\\
    &+C_4e^{-ik\tau-i2\sqrt{2}h\log{|k\tau|}},
\end{align}
where $C_1\sim C_4$ are constants of integration.

\subsubsection{Dynamics of vector modes}
Next we turn to the vector modes. In the single-field inflation, since
the vector modes of metric do not have a source, the vector modes  decay rapidly due to the exponential expansion. 
However,  since we are considering gauge fields, the situation is different. 
After fixing the gauge,  we have two dynamical vector modes $(\mathbb{M}_a)$ (Appendix \ref{quadratic_action}). 
Defining the canonical  variables
\begin{equation}\label{nv1v}
    \boldsymbol{\Delta}_{a}\equiv f\vec{\partial}\mathbb{M}_a .
\end{equation}
Taking the slow roll approximation $\epsilon_E\ll 1$ and using (\ref{ff}),
we obtain the quadratic action as (see Appendix \ref{2action_sl})
\begin{align}\label{vqd}
    S^{(2)}_{\text{vector}}=\frac{1}{2}\int d\tau d^3x &\Bigg\{\boldsymbol{\Delta}_{a}'\cdot  \boldsymbol{\Delta}_{a}'-\partial_i \boldsymbol{\Delta}_{a}\cdot\partial_i \boldsymbol{\Delta}_{a}+\frac{2}{\tau^2}\boldsymbol{\Delta}_{a}\cdot  \boldsymbol{\Delta}_{a}\Bigg\}.
\end{align}
This is the same action as that for a massless field in the de Sitter background 
(here we have three such massless field $a=1,2,3$). 
So we have the equation of motion of these vector modes
\begin{equation}
    \boldsymbol{\Delta}_{a}''+\left(k^2-\frac{2}{\tau^2}\right)\boldsymbol{\Delta}_{a}=0
\end{equation}
On super-horizon scales, the vector modes $\boldsymbol{\Delta}_{a}$ have growing modes, i.e., $\boldsymbol{\Delta}_{a}\sim 1/(-\tau)$, just like as the massless scalar fields. However, the vector modes do not contribute to the curvature perturbations, hence we will not consider them hereafter.

\subsubsection{Dynamics of tensor modes}
Now we turn to the tensor modes. Here, the main aim is to find the enhancement of gravitational waves stemming from the  tensor modes from gauge fields. We also define canonical variables of the tensor modes
\begin{equation}\label{nv1t}
    h_{ij}=aw_{ij},\ \ \ \ \ \ \ t_{ij}=2f\mathbb{T}_{ij}.
\end{equation}
One can expand the tensor modes $h_{ij}$ and $t_{ij}$ with a particular momentum $\bf{k}$ by
the polarization tensors $e^s_{ij}(\hat{\bf{k}})$ as
\begin{equation}
    h_{ij}=\sum_{s=+,\times}e^s_{ij}(\hat{\bf{k}})h_s(\tau), \ \ \ \ \ \ t_{ij}=\sum_{s=+,\times}e^s_{ij}(\hat{\bf{k}})t_s(\tau),
\end{equation}
where $s=+, \times$ represent two polarization tensors  satisfying the normalization relation $e^s_{ij}(\hat{\bf{k}})e^{s'}_{ij}(-\hat{\bf{k}})=\delta^{ss'}$. Now we consider the leading order of slow roll approximation $\epsilon_E\ll 1$ and the de Sitter background $1/\tau=-aH$, the quadratic action can be written as (see Appendix \ref{2action_sl})
\begin{align}
    S^{(2)}_{\text{tensor}}=\frac{1}{8}\sum_{s=+,\times}\int d\tau d^3x &\Bigg\{(h_s')^2-(\partial h_s)^2+(t_s')^2-(\partial t_s)^2+\frac{2}{\tau^2}(h_s)^2+\frac{2}{\tau^2}(t_s)^2\nonumber\\
    &-\frac{8hL_1}{\tau^2}h_st_s+\frac{4hL_1}{\tau}h_st_s'\Bigg\},\label{1ft}
\end{align}
where we have used (\ref{ff}) and replaced $\sqrt{\epsilon_E}$ by $hL_1$. One should note we have assumed $\sqrt{\epsilon_E}=hL_1\ll 1$. The difference with scalar modes is that the interactions between metric and gauge fields is weak $\sim hL_1\ll h$. Because we are working at the leading order of slow roll approximation, both $h$ and $L_1$ can be regarded as constants when $h>\mathcal{O}(1)$. Taking the variation of the action 
with respect to $h_s$ and $t_s$ yields
\begin{align}
    &h_s''+\left(k^2-\frac{2}{\tau^2}\right)h_s+\frac{4hL_1}{\tau^2}t_s-\frac{2hL_1}{\tau}t_s'=0,\\
    &t_s''+\left(k^2-\frac{2}{\tau^2}\right)t_s+\frac{2hL_1}{\tau^2}h_s+\frac{2}{\tau^2}h_s'=0,
\end{align}
where we have replaced $\partial^2\rightarrow -k^2$. We are interested in the dynamics of the modes on superhorizon scales. In this regime, the equations can be approximately solved as
\begin{align}
    h_s=&\frac{\bar{c}_1}{(-\tau)}+\frac{3-\sqrt{9-16h^2L_1^2}}{4hL_1}\bar{c}_3(-\tau)^{(3+\sqrt{9-16h^2L_1^2})/2}\nonumber\label{hs}\\
    &+\frac{3+\sqrt{9-16h^2L_1^2}}{4hL_1}\bar{c}_4(-\tau)^{(3-\sqrt{9-16h^2L_1^2})/2},\\
    t_s=&\bar{c}_2(-\tau)^2+\bar{c}_3(-\tau)^{(3+\sqrt{9-16h^2L_1^2})/2}+\bar{c}_4(-\tau)^{(3-\sqrt{9-16h^2L_1^2})/2},\label{ts}
\end{align}
where $\bar{c}_1\sim\bar{c}_4$ are constants of integration.

\subsection{Two-form gauge fields}
  We consider the scalar, vector and tensor perturbations of the two-form gauge fields. The details of the helicity decomposition can be found in Appendix \ref{2quadratic_action}. There are nine scalar perturbations $(A,B,\delta\phi,\delta\mathbb{B}, \mathbb{X},\mathbb{Z},\mathbb{W},\mathbb{D},\mathbb{V})$, four vector perturbations $(\mathbb{X}_a, \mathbb{Z}_a, \mathbb{V}_a, \mathbb{W}_a)$ and three tensor perturbations $(h_{ij},\mathbb{R}_{ai},\mathbb{S}_{ai})$ in this model. In these perturbations,  $(\delta\mathbb{B}, \mathbb{X},\mathbb{Z},\mathbb{W},\mathbb{D},\mathbb{V})$, ($\mathbb{W}_a$) and $(\mathbb{S}_{ai},\mathbb{R}_{ai})$ are from two-form gauge fields. We fix the gauge of two-form by setting $\mathbb{W}=\mathbb{V}=\mathbb{Z}=\mathbb{X}_a=\mathbb{Z}_a=\mathbb{V}_a=\mathbb{S}_{ai}=0$. Since $(A,B,\mathbb{D},\mathbb{X})$ and $\mathbb{R}_{ai}$ are non-dynamical, those can be eliminated from the quadratic action. The rest dynamical perturbations are $(\delta\phi,\delta\mathbb{B},\mathbb{W}_a)$.

\subsubsection{Dynamics of scalar modes}
Previously, we defined the perturbations of two-form gauge fields $\delta P\equiv g\delta \mathbb{B}/a^2$. Here we also  introduce the canonical variable of two-form gauge fields
\begin{equation}\label{nv2}
    \Delta_{P}\equiv\frac{g}{a}\delta\mathbb{B}.
\end{equation}
The quadratic action of combining scalar parts of gravity, scalar field and gauge field in the slow roll approximation is given by (see Appendix \ref{2action_sl})
\begin{align}
    S^{(2)}_{\text{scalar}}=\frac{1}{2}\int d\tau d^3x& \Bigg\{(\Delta_{\phi}')^2-(\partial\Delta_{\phi})^2+(\Delta_{P}')^2-(\partial\Delta_{P})^2+\frac{2-2l^2}{\tau^2}(\Delta_{\phi})^2\nonumber\\
    &+\frac{2}{\tau^2}(\Delta_P)^2+\frac{8l}{\tau^2}\Delta_{\phi}\Delta_P-\frac{4l}{\tau}\Delta_{\phi}\Delta_P'\Bigg\},\label{2action2per}
\end{align}
where $l$ is given by (\ref{h}) and can be regarded as a constant at the leading order in the slow roll approximation. Taking the variation of the action with respect to $\Delta_{\phi}$ and $\Delta_p$ gives
\begin{align}
    &\Delta_{\phi}''+\left(k^2-\frac{2-2l^2}{\tau^2}\right)\Delta_{\phi}-\frac{4l}{\tau^2}\Delta_P+\frac{2l}{\tau}\Delta_P'=0,\label{DeltaPhieq2}\\
    &\Delta_P''+\left(k^2-\frac{2}{\tau^2}\right)\Delta_P-\frac{2l}{\tau^2}\Delta_{\phi}-\frac{2l}{\tau}\Delta_{\phi}'=0.\label{DeltaPeq}
\end{align}
Firstly on superhorizon scales ($|k\tau|\ll 1$), the solutions of these equations are approximately given by
\begin{align}
    \Delta_{\phi}=&\frac{3}{l}\frac{d_1}{(-\tau)}+\frac{\sqrt{9-24l^2}-3}{4l}d_3(-\tau)^{(1+\sqrt{9-24l^2})/2}\nonumber\\
    &-\frac{\sqrt{9-24l^2}+3}{4l}d_4(-\tau)^{(1-\sqrt{9-24l^2})/2},\\
    \Delta_P=&\frac{d_1}{(-\tau)}+d_2(-\tau)^2+d_3(-\tau)^{(1+\sqrt{9-24l^2})/2}+d_4(-\tau)^{(1-\sqrt{9-24l^2})/2},
\end{align}
where $d_1\sim d_4$ are constants of integration. Again, $d_2$ modes decay and become irrelevant soon. For $l>\sqrt{9/24}\simeq0.6$, $d_3$ and $d_4$ also decay rapidly. Only $d_1$ is the growing mode that contributes to the curvature perturbations. 

In order to determine the amplitude of the growing mode, we need to known the sub-horizon evolution. 
The solutions on sub-horizon scales are
\begin{align}
    \Delta_{\phi}=&iD_1 e^{ik\tau+il\log{|k\tau|}}-iD_2e^{ik\tau-il\log{|k\tau|}}+iD_3e^{-ik\tau-il\log{|k\tau|}}\nonumber\\
    &-iD_4e^{-ik\tau-il\log{|k\tau|}},\\
    \Delta_{P}=&D_1e^{ik\tau+il\log{|k\tau|}}+D_2e^{ik\tau-il\log{|k\tau|}}+D_3e^{-ik\tau+il\log{|k\tau|}}\nonumber\\
    &+D_4e^{-ik\tau-il\log{|k\tau|}},
\end{align}
where $D_1\sim D_4$ are constants of integration. Both the one-form and two-form cases have a correction depending on $l\log{|k\tau|}$ in the above solution. They do not matter in the very early time of inflation ($k\tau=-\infty$) but become important around  the horizon crossing.

\subsubsection{Dynamics of vector modes}
In the case of the two-form gauge fields, after fixing the gauge and eliminating the non-dynamical degrees of freedom, the  dynamical vector modes $(\mathbb{W}_a)$ remain (see Appendix \ref{2quadratic_action}). 
Defining the canonical vector variables
\begin{equation}\label{nv2v}
    \boldsymbol{\Delta}_{a}\equiv \frac{g}{a}\vec{\partial}\mathbb{W}_a \ ,
\end{equation}
and using the slow roll approximation $\epsilon_H,\epsilon_B\ll 1$ and (\ref{gg}), we can reduce the quadratic action to the same form as (\ref{vqd}) (see Appendix \ref{2action_sl}). Like the one-form case, $\boldsymbol{\Delta}_{a}$ have a growing mode $\boldsymbol{\Delta}_{a}\sim 1/(-\tau)$. However, since they do not contribute to curvature perturbations,  we do not discuss them hereafter.

\subsubsection{Dynamics of tensor modes}
In contrast to the one-form gauge fields, there is no dynamical tensor degree of freedom of the two-form gauge fields. The only physical modes we need to treat are metric perturbations $w_{ij}$. After eliminating the non-dynamical modes (see Appendix \ref{2quadratic_action}) and defining the canonical variables
\begin{equation}\label{nv2t}
    h_{ij}\equiv aw_{ij} ,
\end{equation}
In the leading order of the slow roll approximation $\epsilon_B\ll 1$
\begin{align}
    S^{(2)}_{\text{tensor}}=\frac{1}{8}\int d\tau d^3x \Bigg\{(h_{ij}')^2-(\partial h_{ij})^2+\frac{a''}{a}(h_{ij})^2
    \Bigg\}.
\end{align}
We can see the quadratic action is the same as that of the single-field inflation (see Appendix \ref{2action_sl}). That is, for the two-form gauge fields, the power spectrum of tensor perturbations is not modified. This is because there is no source of gravitational waves from two-form fields.

\subsection{Numerical analysis}
To understand the behaviors of these modes across the entire scales, in this subsection we  study the solutions numerically. We compute the one-form case and two-form case separately. We can introduce dimensionaless quantities $\tilde{k}\equiv k/k_0$, $\tilde{\tau}\equiv k_0\tau$, $\tilde{\Delta}_{\phi}\equiv\sqrt{k_0}\Delta_{\phi}$, $\tilde{\Delta}_{Q}\equiv\sqrt{k_0}\Delta_{Q}$ and $\tilde{\Delta}_{P}\equiv\sqrt{k_0}\Delta_{P}$, where $k_0$ is a free parameter and we set $\tilde{k}=10^{-3}$. The coefficients of sub-horizon modes $C_i$ and $D_i$ can be fixed by the initial conditions of the quantum state, i.e., the Bunch-Davies vacuum in de Sitter space. Similarly to the hyperbolic inflation, we can fixed them as: $C_1=C_2=C_3=D_1=D_2=D_3=0$ and $C_4=D_4=1/\sqrt{2\tilde{k}}$ for $\tilde{\Delta}_{\phi}$, $\tilde{\Delta}_{Q}$ and $\tilde{\Delta}_{P}$, and then solve the equations of motion (\ref{DeltaPhieq}), (\ref{DeltaQeq}) and (\ref{DeltaPhieq2}), (\ref{DeltaPeq})  with initial conditions
\begin{equation}
    \tilde{\Delta}_{\phi}(\tau_0)=\frac{-i}{\sqrt{2\tilde{k}}}\left(1-\frac{i}{\tilde{k}\tilde{\tau}_0}\right)e^{-i\tilde{k}\tilde{\tau_0}},\ \ \ \ \      \tilde{\Delta}_{Q}(\tau_0)=\tilde{\Delta}_{P}(\tau_0)=\frac{1}{\sqrt{2\tilde{k}}}\left(1-\frac{i}{\tilde{k}\tilde{\tau}_0}\right)e^{-i\tilde{k}\tilde{\tau_0}} \ .
\end{equation}
deeply inside the horizon $\tilde{\tau}_0=10^{-7}$.
\begin{figure}[tbp]
\centering
\includegraphics[scale=0.62]{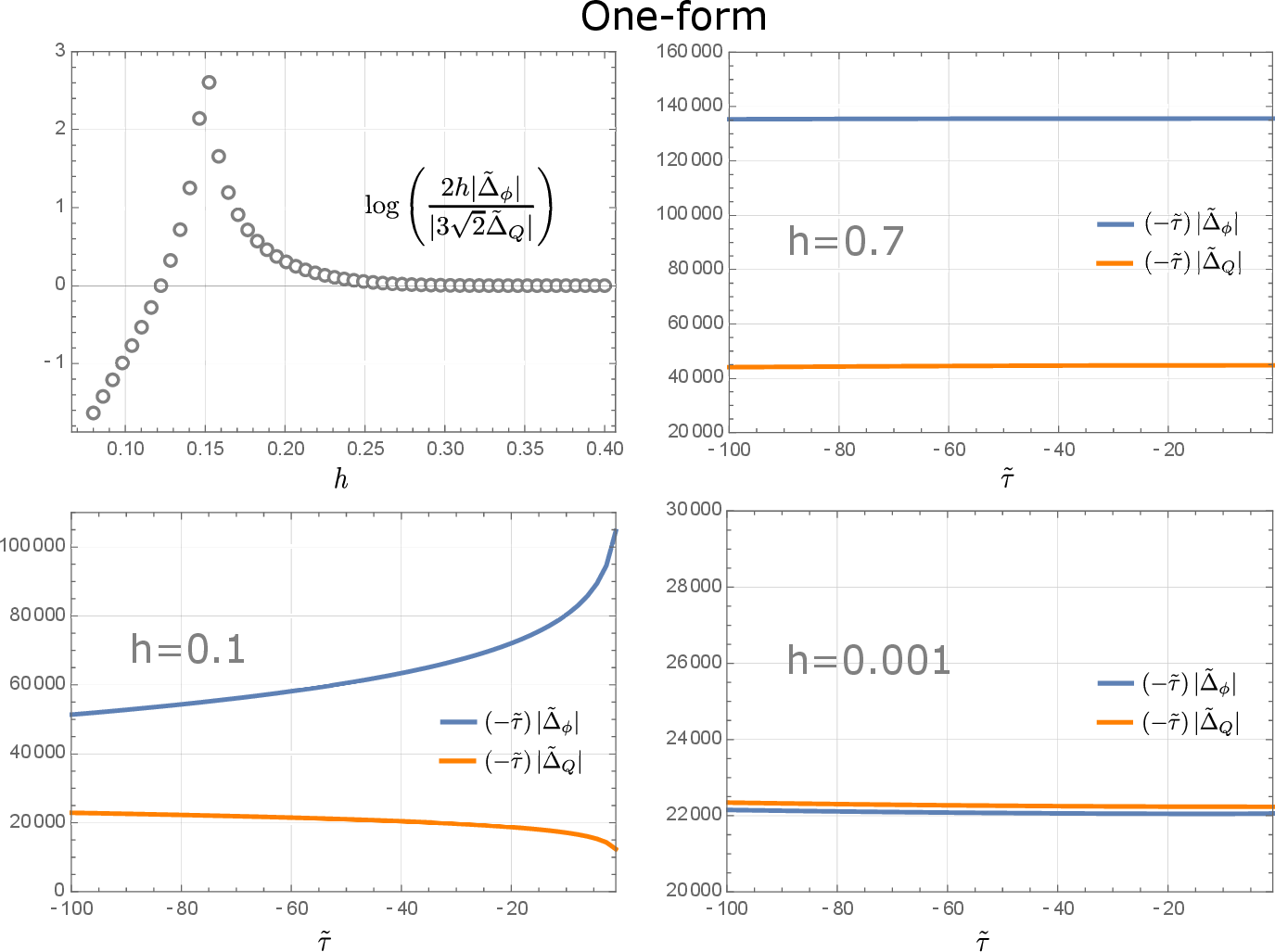}
\caption{\label{fig:deltaG}($Top-Left$) The function $\log{(2h(|\tilde{\Delta}_{\phi}|/|\tilde{\Delta}_Q|)/(3\sqrt{2}))}$ for various values of $h$. It remains zero until $h\lesssim\sqrt{9/96}\simeq 0.3$. ($Top-Right$) The evolution of $(-\tilde{\tau})\tilde{\Delta}_{\phi}$ and $(-\tilde{\tau})\tilde{\Delta}_{Q}$ on super-horizon scale for $h=0.7$. Both of them are almost constants. ($Bottom-Left$) The evolution of $(-\tilde{\tau})\tilde{\Delta}_{\phi}$ and $(-\tilde{\tau})\tilde{\Delta}_{Q}$ on super-horizon scale for $h=0.1$. They deviate from constants because of the slow decay of $c_4$ modes. ($Bottom-Right$) The evolution of $(-\tilde{\tau})\tilde{\Delta}_{\phi}$ and $(-\tilde{\tau})\tilde{\Delta}_{Q}$ on super-horizon scales for $h=0.001$. Both of them are almost constants because $c_4$ modes are also nearly scale invariant. ($M_{\text{pl}}=1$, $\tilde{k}=10^{-3}$)} 
\end{figure}
\begin{figure}[tbp]
\centering
\includegraphics[scale=0.62]{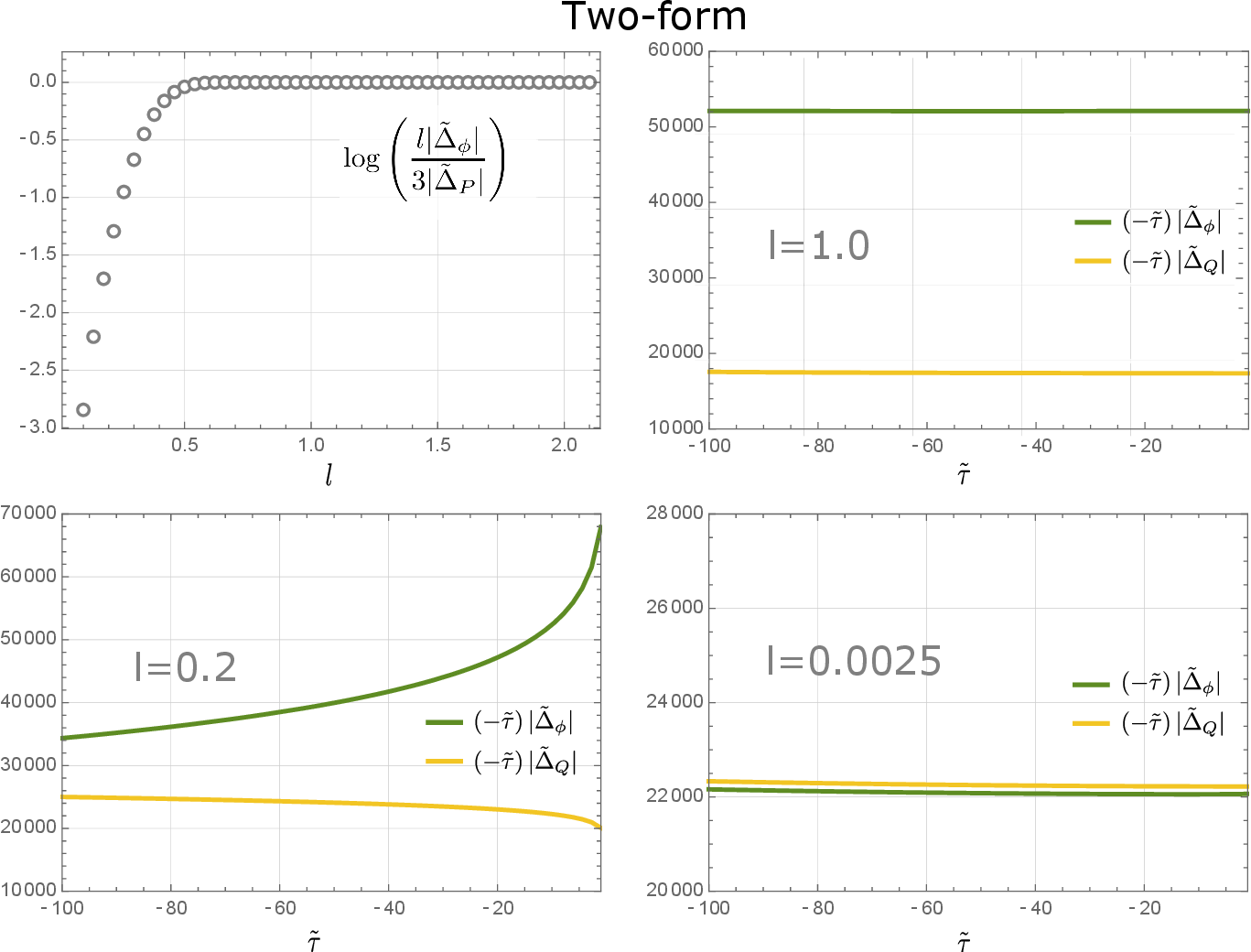}
\caption{\label{fig:2deltaG}($Top-Left$) The function $\log{(l(|\tilde{\Delta}_{\phi}|/|\tilde{\Delta}_P|)/3)}$ for various values of $l$. It remains zero until $l\lesssim\sqrt{9/24}\simeq 0.6$. ($Top-Right$) The evolution of $(-\tilde{\tau})\tilde{\Delta}_{\phi}$ and $(-\tilde{\tau})\tilde{\Delta}_{P}$ on super-horizon scale for $l=1.0$. Both of them are almost constants. ($Bottom-Left$) The evolution of $(-\tilde{\tau})\tilde{\Delta}_{\phi}$ and $(-\tilde{\tau})\tilde{\Delta}_{P}$ on super-horizon scale for $l=0.2$. They deviate from constants because of the slow decay of $d_4$ modes. ($Bottom-Right$) The evolution of $(-\tilde{\tau})\tilde{\Delta}_{\phi}$ and $(-\tilde{\tau})\tilde{\Delta}_{P}$ on super-horizon scales for $h=0.0025$. Both of them are almost constants because $d_4$ modes are also nearly scale invariant. ($M_{\text{pl}}=1$, $\tilde{k}=10^{-3}$)} 
\end{figure}

We first consider the super-horizon behaviours. However, we should divide the behaviours of modes on super-horizon scales into three regimes depending on $h$ for one-form case and $l$ for two-form case : $h\ll1$, $h\lesssim\sqrt{9/96}\simeq0.3$ and $h>0.3$ for one-form case and $l\ll1$, $l\lesssim\sqrt{9/24}\simeq0.6$ and $l>0.6$ for two-form case. 
\begin{itemize}
\item[(1)] When $h>0.3$($l>0.6$), the $c_3$ and $c_4$($d_3$ and $d_4$) modes are massive modes and decay away rapidly. Then the only remaining modes are adiabatic modes $c_1$($d_1$). In Figure \ref{fig:deltaG} ($Top-Right$)(Figure \ref{fig:2deltaG}($Top-Right$)) we choose $h=0.7$($l=1.0$), which corresponds to sizable one-form(two-form) gauge fields. Then we plot the evolution of modes $\tilde{\Delta}_{\phi}$ and $\tilde{\Delta}_{Q}$($\tilde{\Delta}_{P}$) on super-horizon scales. We see $\tilde{\Delta}_{\phi}$ and $\tilde{\Delta}_{Q}$($\tilde{\Delta}_{P}$) behave as growing modes so that $\delta\phi$ and $\delta Q$($\delta P$) are almost constant modes and have ratio $|\tilde{\Delta}_{\phi}/\tilde{\Delta}_{Q}|=3\sqrt{2}/(2h)$($|\tilde{\Delta}_{\phi}/\tilde{\Delta}_{P}|=3/l$) on large scales ($\tilde{\tau}> -100$), as we expected. 

\item[(2)] When $h\lesssim0.3$($l\lesssim0.6$), the $c_3$ and $c_4$($d_3$ and $d_4$) modes are also decaying modes. But they decay slowly and are comparable to the adiabatic $c_1$($d_1$) modes during inflation. Hence, we can not ignore them. Then ratio $(2h/(3\sqrt{2}))|\tilde{\Delta}_{\phi}/\tilde{\Delta}_{Q}|$($(l/3)|\tilde{\Delta}_{\phi}/\tilde{\Delta}_{P}|$) is dependent on $h$($l$), see Figure \ref{fig:deltaG} ($Top-Left$) and Figure \ref{fig:2deltaG} ($Top-Left$). So $\delta\phi$ and $\delta Q$($\delta P$) deviate from constant as is seen in Figure \ref{fig:deltaG} ($Bottom-Left$) and Figure \ref{fig:2deltaG} ($Bottom-Left$). 

\item[(3)] When $h\ll 1$($l\ll 1$), where the energy density of gauge fields is small compared to kinetic energy of $\phi$, the $c_4$($d_4$) modes are also comparable to $c_1$($d_1$) modes. Nevertheless, $(1-\sqrt{9-96h^2})/2\simeq-1$($(1-\sqrt{9-24h^2})/2\simeq-1$) hence $c_4$($d_4$) modes behave 
like the adiabatic growing modes $c_1$($d_1$). In this case, $\delta\phi$ and $\delta Q$($\delta P$) are also nearly constant modes on super-horizon scales, see Figure \ref{fig:deltaG} ($Bottom-Right$) and Figure \ref{fig:2deltaG} ($Bottom-Right$).
\end{itemize}

\begin{figure}[tbp]
\centering
\includegraphics[scale=0.465]{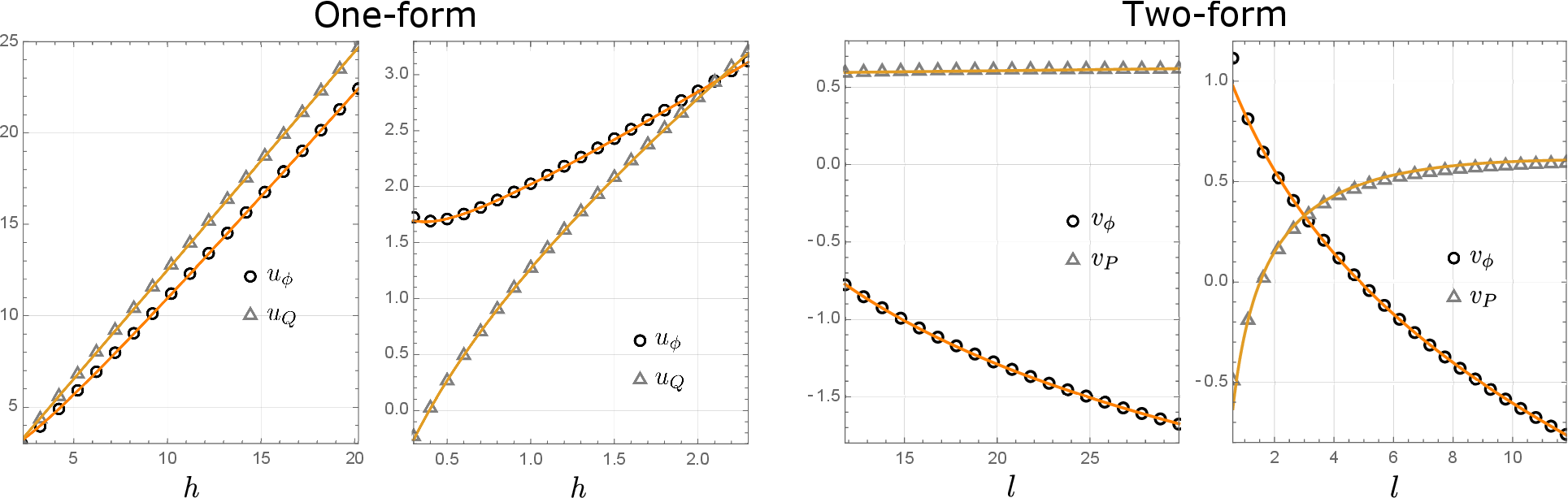}
\caption{\label{fig:g-hl}($Left$) The function $u_i\equiv\log{(|\tilde{\Delta}_i(h, \tilde{\tau}_{\text{late}})|/|\tilde{\Delta}^0_i(\tilde{\tau}_{\text{late}})|)}$ of one-form case for various values of $h\in[2.3,20.2]$ and $h\in[0.3,2.3]$. They can be nearly fitted by functions (\ref{gfunction}) and (\ref{gfunction2}) respectively. ($Right$) The function $v_i\equiv\log{(|\tilde{\Delta}_i(l, \tilde{\tau}_{\text{late}})|/|\tilde{\Delta}^0_i(\tilde{\tau}_{\text{late}})|)}$ of two-form case for various values of $l\in[11.8,29.8]$ and $l\in[0.6,11.8]$. They can be nearly fitted by functions (\ref{2gfunction}) and (\ref{2gfunction2}) respectively. ($M_{\text{pl}}=1$, $\tilde{k}=10^{-3}$, $\tilde{\tau}_{\text{late}}=-1$)} 
\end{figure}

To determine the coefficient $c_i$ and $d_i$ on super-horizon scales, we need to match sub-horizon solutions with the super-horizon solution at the horizon crossing. Let us first consider the one-form case. In contrast to the single-field inflation, the mapping from sub-horizon mode with $C_4$ to super-horizon mode with $c_1$ gives the exponentially large amplitude, To see this enhancement of amplitude, we note that from the quadratic action (\ref{action2per}), we can obtain the eigenvalues $m^2_{\pm}=(k^2\tau^2-2-4h^2\pm4h\sqrt{8+h^2})(1/\tau^2)$ by diagonalizing the mass matrix. The mass squared becomes tachyonic for $k^2\tau^2\lesssim 2+4h^2\pm4h\sqrt{8+h^2}$ so perturbations are enhanced depending on the value of $h$. To find this $h$ dependence, we numerically study the behavior of perturbations $\tilde{\Delta}_i(h)$ compared to $\tilde{\Delta}^0_i\equiv\tilde{\Delta}_i(h=0)$ on super-horizon scales, where $i=\phi,Q$. One can define the function of $h$ as \cite{Mizuno:2017idt}
\begin{equation}
    u_i(h)\equiv \log{\frac{|\tilde{\Delta}_i(h, \tilde{\tau}_{\text{late}})|}{|\tilde{\Delta}^0_i(\tilde{\tau}_{\text{late}})|}},
\end{equation}
where $\tilde{\tau}_{\text{late}}$ is the late time where perturbations $(-\tilde{\tau})\tilde{\Delta}_i$ become constant and we set $\tilde{\tau}_{\text{late}}=-1$. We plot function $u_i(h)$ for various values of $h$ in Figure \ref{fig:g-hl} ($left$) . The dependence of $h$ can be roughly divided into two parts: $0.3<h< 2.3$ and $h>2.3$. First, in the $h> 2.3$ regime, we find the dependence of $u_{Q}$ on $h$ is nearly linear (Figure \ref{fig:g-hl} ($Left$)). Then we can use a linear function to approximately fit the function. Moreover, from the relation of coefficients of adiabatic mode between $\tilde{\Delta}_{\phi}$ (\ref{superphi}) and $\tilde{\Delta}_Q$ (\ref{superq}), we can also obtain the function of $u_{\phi}$. In conclusion, we have
\begin{equation}
   \ \ \ \ \ \ \ \ \ u_{\phi}\simeq q_1+p_1h+\log{\left(3\sqrt{2}/(2h)\right)},\ \ \ \ \ \ \ \ \ \ u_{Q}\simeq q_1+p_1h\ \ \ \ \ \ \ \ \ (h>2.3),\label{gfunction}
\end{equation}
where $p_1\simeq1.193$ and $q_1\simeq0.580$. In other words, perturbations $\tilde{\Delta}_{\phi}$ and $\tilde{\Delta}_{Q}$ have exponentially large enhancements of amplitude in $h$. On the other hand, for the $h\lesssim\mathcal{O}(1)$ regime, the situation becomes more complicated (Figure \ref{fig:g-hl} ($Right$)). We also roughly fit the curve of $h\in[0.3,2.3]$ by
\begin{equation}
\begin{aligned}
    \ \ \ \ \ \ \ \ \ \ \ &u_{\phi}\simeq q_2+p_2h+\log{\left(m h+n\right)}+\log{\left(3\sqrt{2}/(2h)\right)}\\ 
    &u_{Q}\simeq q_2+p_2h+\log{\left(m h+n\right)}\ \ \ \ \ \ \ \ \ 
\end{aligned},\ \ \ \ \ \ \ \ \ (0.3\lesssim h\lesssim 2.3)\label{gfunction2}
\end{equation}
where $p_2\sim0.910$, $q_2\simeq0.197$, $m\simeq0.990$ and $n\simeq0.185$. We can also see the exponential enhancements of $\tilde{\Delta}_{\phi}$ and $\tilde{\Delta}_{Q}$.

The two-form case is similar. We define the function of $l$
\begin{equation}
    v_i(l)\equiv \log{\frac{|\tilde{\Delta}_i(l, \tilde{\tau}_{\text{late}})|}{|\tilde{\Delta}^0_i(\tilde{\tau}_{\text{late}})|}},
\end{equation}
where $i=\phi,P$ and here $\tilde{\Delta}^0_i\equiv \tilde{\Delta}_i(l=0)$. The eigenvalues of mass in the action (\ref{2action2per}) are $m^2_{\pm}=(k^2\tau^2-2+l^2\pm l\sqrt{16+l^2})(1/\tau^2)$. Hence the mass squared becomes tachyonic when $k^2\tau^2\lesssim 2-l^2\pm l\sqrt{16+l^2}$. We also plot the function $v_i(l)$ for $\tilde{\Delta}_{\phi}$ and $\tilde{\Delta_{P}}$ in Figure \ref{fig:g-hl} ($Right$). We divide the function into two parts: $0.6<l<11.8$ and the linear part $l>11.8$. For the linear part, we use the functions
\begin{equation}
    \ \ \ \ \ \ \ \ \ v_{\phi}\simeq s_1+r_1l+\log{\left(3/l\right)}, \ \ \ \ \ \ \ \ \ \ \ v_{P}\simeq s_1+r_1l\ \ \ \ \ \ \ \ \ \ \ (l>11.8),\label{2gfunction}
\end{equation}
to fit them, where $r_1=0.00136$ and $s_1=0.580$. We find the growth of the function $v_{P}$ is very slow even when $l\gg1$, which means that when crossing the horizon, the enhancement of $\Delta_{P}$ is not significant even when the energy density of two-form gauge fields is very large. This is because, for large $l$, the eigenvalue of mass in action $m^2_{+}$ becomes tachyonic when $k^2\tau^2\lesssim 2-l^2+l\sqrt{16+l^2}\simeq10$, which is independent of $l$. That is, the enhancement of mode with momentum $k$ always start at $|k\tau|\simeq \sqrt{10}$ and end when crossing the horizon. So the enhancement does not increase any more if $l$ is large enough. On the other hand, we have $v_{\phi}<0$ for large $l$, which means that the $\Delta_{\phi}$ mode is exponentially suppressed. These are the difference from the one-form case. For $0.6<l<11.8$, we also use functions
\begin{equation}
\begin{aligned}
    \ \ \ \ \ \ \ \ \ \ \ &v_{\phi}\simeq s_2+r_2l-\log{\left(m' l+n'\right)}\\ 
    &v_{P}\simeq s_2+r_2l-\log{\left(m' l+n'\right)}-\log{\left(3/l\right)}
\end{aligned},\ \ \ \ \ \ \ \ \ (0.6\lesssim l\lesssim 11.8)\label{2gfunction2}
\end{equation}
to approximately fit them, where $r_2=0.0115$, $s_2=0.695$, $m'=0.268$ and $n'=0.588$. The growth of linear part is slow ($r_2\simeq0.01$) while the negative logarithmic part becomes large enough. Hence $v_{\phi}$ becomes negative.

\begin{figure}[tbp]
\centering
\includegraphics[scale=0.57]{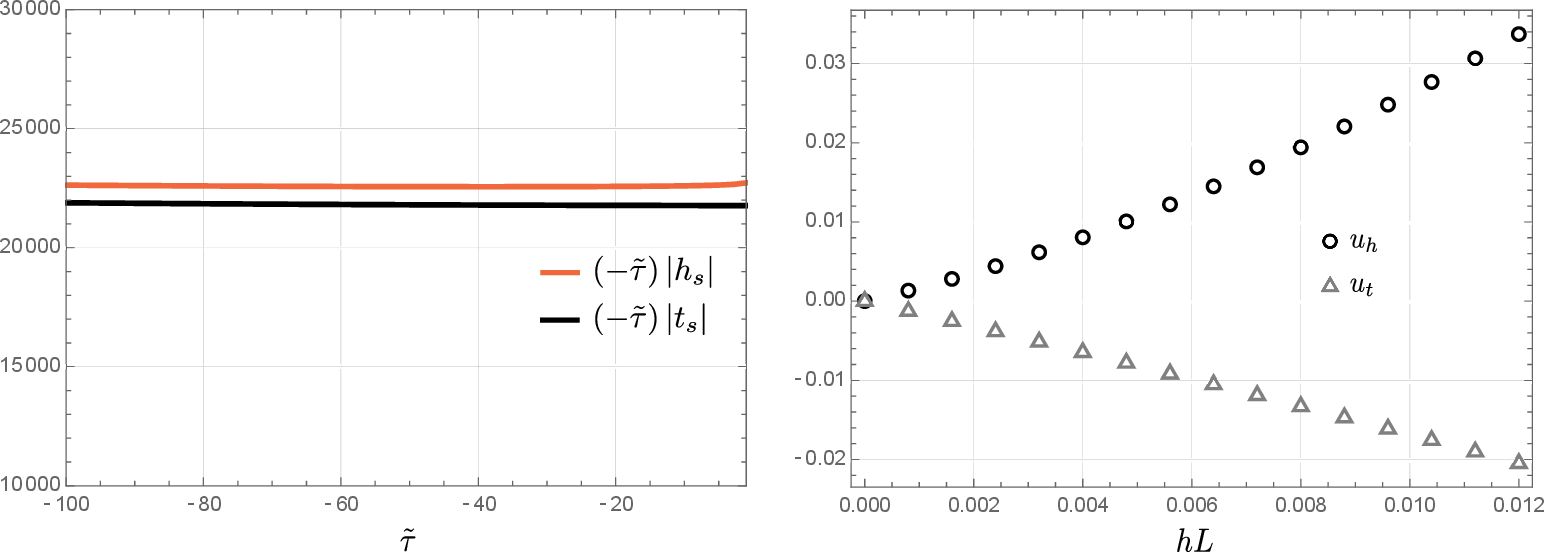}
\caption{\label{fig:GW}($Left$) The evolution of $(-\tilde{\tau})\tilde{h}_s$ and $(-\tilde{\tau})\tilde{t}_s$ on super-horizon scale for $hL_1=0.01$. Both of them are almost constants because $\bar{c}_4$ modes are also nearly scale invariant. ($Right$) The function $u_h\equiv\log{(|\tilde{h}_s(h, \tilde{\tau}_{\text{late}})|/|\tilde{h}^0_s(\tilde{\tau}_{\text{late}})|)}$ for various values of $hL\in[0,0.012]$. ($M_{\text{pl}}=1$, $\tilde{k}=10^{-3}$, $\tilde{\tau}_{\text{late}}=-1$)} 
\end{figure}
Now let us see the tensor modes of the one-form case. In (\ref{ts}), the $\bar{c}_2$ mode rapidly becomes irrelevant.
Hence we can ignore it.  Because $hL_1\ll1$, $\bar{c}_3$ modes will decay soon. For gravitational wave $h_s$, there are two contributions, $\bar{c}_1$ and $\bar{c}_4$ modes. The $c_1$ mode is the growing adiabatic mode while $\bar{c}_4$ is a decaying mode. However, we have $(3-\sqrt{9-16h^2L_1^2})/2\simeq -1$ for $hL_1\ll1$, which means they decay very slowly. Thus, $\tau\bar{c}_4$ can be regraded as nearly constant modes. This is the same for $t_{ij}$ in which only $\bar{c}_4$ modes left. The evolution of $(-\tau)h_s$ and $(-\tau)t_s$ are shown in Figure \ref{fig:GW} ($Left$). 

We can also use the free parameter $k_0$ to normalize the modes as $\tilde{h}_s\equiv\sqrt{k_0}h_s$ and $\tilde{t}_s\equiv\sqrt{k_0}t_s$. We can define the function by comparing $\tilde{h}_s(h)$ with $\tilde{h}^0_s\equiv\tilde{h}_s(h=0)$ as
\begin{equation}
    u_h(h)\equiv\log{\frac{|\tilde{h}_s(h,\tilde{\tau}_{\text{late}})|}{|\tilde{h}^0_s(\tilde{\tau}_{\text{late}})|}},
\end{equation}
where $\tilde{\tau}_{\text{late}}=-1$ was chosen. We plot the digram $u_h$-$hL$ in Figure \ref{fig:GW} ($Right$) in the interval $0<hL\lesssim0.01$. We find in this regime $u_h\sim\mathcal{O}(hL)$. In other words, the enhancement owing to gauge fields is sub-leading compared to the amplitude of tensor mode itself.

\section{Primordial power spectrum}\label{PPS}
Until now, we have investigated the behaviors of scalar perturbations in regime $h,l>\mathcal{O}(1)$ and tensor perturbations in regime $hL_1, lL_2\ll1$, whose amplitudes are conserved on super-horizon scales. We can now discuss the power spectrum of this model and compare it with observations. Here we also only consider the regime that $h,l>\mathcal{O}(1)$ and $hL_1,lL_2\ll 1$. 
Firstly we should note in the previous section we introduced a free scale $k_0$ to normalize the quantities with dimensions. We can define the corresponding dimensionaless power spectrum,
\begin{equation}
    \mathcal{P}_{\tilde{\Delta}_{\phi}}\equiv\frac{\tilde{k}^3}{2\pi^2}|\tilde{\Delta}_{\phi}|^2,\ \ \ \ \ \ \mathcal{P}_{\tilde{\Delta}_{Q}}\equiv\frac{\tilde{k}^3}{2\pi^2}|\tilde{\Delta}_{Q}|^2,\ \ \ \ \ \ \mathcal{P}_{\tilde{\Delta}_{P}}\equiv\frac{\tilde{k}^3}{2\pi^2}|\tilde{\Delta}_{P}|^2.
\end{equation}
We see that $\tilde{\tau}^2\mathcal{P}_{\tilde{\Delta}_{i}}=\tau^2\mathcal{P}_{\Delta_{i}}$ holds for any scale $k_0$ we choose. In Figure \ref{fig:PowerSpectrum} we plot the power spectrum of these modes multiplied by $\tilde{\tau}^2$ against the wavenumber in the range $0.0001<k<0.001$. We can see they are almost scale invariant. We will use them to calculate the power spectrum of curvature perturbations.

\subsection{One-form gauge fields}
\begin{figure}[tbp]
\centering
\includegraphics[scale=0.66]{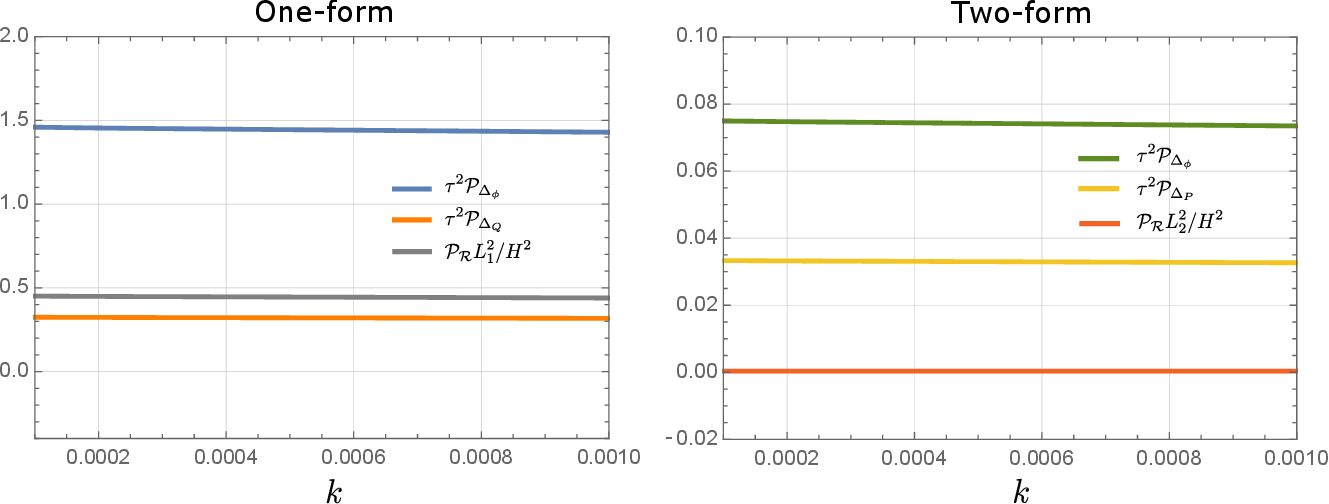}
\caption{\label{fig:PowerSpectrum}($Left$) The power spectrum  $\tilde{\tau}^2\mathcal{P}_{\tilde{\Delta}_{\phi}}$ (blue), $\tilde{\tau}^2\mathcal{P}_{\tilde{\Delta}_{Q}}$ (orange) and $\mathcal{P}_{\mathcal{R}}L^2/H^2$ (gray) for one-form case on super-horizon scale when $h=1.0$. ($Right$) The power spectrum  $\tilde{\tau}^2\mathcal{P}_{\tilde{\Delta}_{\phi}}$ (green), $\tilde{\tau}^2\mathcal{P}_{\tilde{\Delta}_{P}}$ (yellow) and $\mathcal{P}_{\mathcal{R}}L^2/H^2$ (red) for two- form case on super-horizon scale at $l=2.0$. All of them are evaluated at $\tilde{\tau}=\tilde{\tau}_{\text{late}}=-1$ and nearly scale invariant. ($M_{\text{pl}}=1$)} 
\end{figure}

In contrast to the single-field inflation, there are two contributions of adiabatic modes to the curvature perturbation, one from the scalar field and the other from the scalar parts of gauge fields. The mode $U_i$ corresponds to the fluctuation of the magnetic, which has vanishing background value so decouples with other modes. In other words, $U$ is a pure isocurvature mode \cite{Gorji:2020vnh}. Hence only the $\delta Q$ contributes to the curvature perturbations. The curvature perturbation in spatially flat gauge $\psi=0$ is defined by
\begin{equation}
    \mathcal{R}\equiv H\delta u,\label{R}
\end{equation}
where $\delta u$ is the velocity potential given by $\delta T^0_i\equiv(\rho+p)\partial_i\delta u$. For the isotropic one-form gauge fields, we have $\delta T^0_i=-\dot{\phi}\partial_i\delta\phi-(2f^2\dot{\mathbb{A}}/a^2)\partial_i\delta\mathbb{A}$. Here $\rho$ and $p$ are the total energy density and pressure of the system and in our model we have $\rho+p=\dot{\phi}^2+2f^2\dot{\mathbb{A}}^2/a^2$. Then we obtain \cite{Firouzjahi:2018wlp,Gorji:2020vnh}
\begin{equation}
    \mathcal{R}=-H\frac{\dot{\phi}\delta\phi+(2f^2\dot{\mathbb{A}}^2/a^2)\delta\mathbb{A}}{\dot{\phi}^2+2f^2\dot{\mathbb{A}}^2/a^2}.
\end{equation}
On the super-horizon scales, for $h>0.3$, the contribution to the $\mathcal{R}$ comes from the adiabatic $c_1$ modes in (\ref{superphi}) and (\ref{superq}) so we have $\Delta_{\phi}/\Delta_{Q}=-3\sqrt{2}/(2h)$. Then we can obtain 
\begin{equation}
    \mathcal{R}=-\frac{1}{aL_1}\frac{-\Delta_{\phi}+\sqrt{2}h\Delta_Q}{1+2h^2}=-\frac{\sqrt{2}}{L_1}\frac{3+2h^2}{2h(1+2h^2)}\frac{\Delta_{Q}}{a},
\end{equation}
where we have used $\dot{\phi}/H=-L_1$ and $h\equiv\sqrt{\epsilon_E}/L_1$. On the other hand, we also need to know the amplitude of the power spectrum. We note for $h=0$, the equation of perturbation $\tilde{\Delta}_{Q}$ is the same as that of a scale field in de Sitter space. Hence we have $\mathcal{P}_{\tilde{\Delta}_{Q}}/a^2(h=0)=H^2/(2\pi)^2$, which implies $\tilde{\tau}^2\mathcal{P}_{\tilde{\Delta}_{Q}}(h=0)=\tau^2\mathcal{P}_{\Delta_{Q}}(h=0)=1/(2\pi^2)$. We have studied the enhancements of amplitude of $\tilde{\Delta}_{Q}$ for $h\simeq {\cal O} (1)$ in (\ref{gfunction}) and (\ref{gfunction2}). Then after using (\ref{h}) and $\tilde{\tau}^2\mathcal{P}_{\tilde{\Delta}_{i}}=\tau^2\mathcal{P}_{\Delta_{i}}$ we finally have
\begin{equation}
    \mathcal{P}_{\mathcal{R}}=\frac{2H^2}{L_1^2}\left(\frac{3+2h^2}{2h(1+2h^2)}\right)^2\tau^2\mathcal{P}_{\Delta_{Q}}=\frac{1}{(2\pi)^2}\frac{H^2}{\epsilon_H}\frac{\left(3+2h^2\right)^2}{4h^2\left(1+2h^2\right)}e^{2u_{Q}(h)},\label{Rps}
\end{equation}
which is scale invariant on super-horizon scales (see Figure \ref{fig:PowerSpectrum} ($Left$)). We can also see the exponential growth $\sim e^{2u_{Q}}$ of the power spectrum. In the second equality, we have used  $\epsilon_H=L_1^2/2+\epsilon_E=L_1^2(1+2h^2)/2$. The power spectrum is a function of energy ratio $h$ and $\epsilon_H$.

We can then use (\ref{Rps}) to calculate the spectral index of $\mathcal{R}$. Using  $\epsilon_H=L_1^2/2+\epsilon_E=L_1^2(1+2h^2)/2$, we can replace the $h$ in (\ref{Rps}) with $\epsilon_H$ and $L_1$. Note that, in contrast to hyperbolic inflation, now $L_1$ is also a slow roll variable and the corresponding slow roll parameter is defined by (\ref{lchang}). 
Modes with different momentum exit the horizon at different times. When the mode with momentum $k$ crosses the horizon $k=aH$ we have $d(\ln{k})\simeq dN$, where $N$ is the e-folding number before the end of inflation. Then we obtain
\begin{equation}
    \frac{dh^2}{d\ln{k}}\simeq\frac{dh^2}{dN}=\frac{1}{2}\left(1+2h^2\right)\left(\eta_H-\eta_L\right).\label{hk}
\end{equation}
Note that in two different regimes $0.3< h\lesssim2.3$ and $h>2.3$ there are two different enhancements functions. However, in general using (\ref{Rps}) the spectral index of curvature can be given by
\begin{align}
    n_s-1&\equiv \frac{d\ln{\mathcal{P}_{\mathcal{R}}}}{d\ln{k}}\simeq\frac{d\ln{\mathcal{P}_{\mathcal{R}}}}{dN}\nonumber\\
    &=-2\epsilon_H-\eta_H+\left(\frac{2h^2-1}{3+2h^2}-\frac{1+2h^2}{2h^2}+\frac{du_Q}{dh}\frac{1+2h^2}{2h}\right)\left(\eta_H-\eta_L\right).
\end{align}
We see the spectral tilt depends on the energy ratio $h$. If we consider $h\gg 1$, i.e., the energy of gauge field is large compared to the kinetic energy of scalar field (but small compared to the potential energy), then  $\epsilon_H\simeq\epsilon_E\gg L_1^2/2$ and we obtain $n_s-1\simeq-2\epsilon_H-\eta_H+p_2h(\eta_H-\eta_L)$. The third term will be dominant unless $\eta_H-\eta_L\ll\epsilon_H$.

Now we calculate the power spectrum of gravitational waves $\mathcal{P}_{T}$. First we also have $\tilde{\tau}^2\mathcal{P}_{\tilde{h}}=\tau^2\mathcal{P}_{h}$ for any sclae $k_0$ and when $h=0$ the amplitude is normalized as $\mathcal{P}_{\tilde{w}}(h=0)=4H^2/(2\pi)^2$, which implies $\tau^2\mathcal{P}_h(h=0)=4/(2\pi)^2$. Therefore the amplitude  $\tau^2\mathcal{P}_h=4\exp{(g_h)}/(2\pi)^2$ when $h\neq0$. Then we have
\begin{equation}
    \mathcal{P}_T=\sum_{s=+,\times}H^2\tau^2\mathcal{P}_h=\frac{8H^2}{(2\pi)^2}e^{2u_h}.
\end{equation}
The spectral index of gravitational waves $n_T$ is given by
\begin{equation}
    n_T\equiv\frac{d\ln{\mathcal{P}_T}}{d\ln{k}}\simeq\frac{d\ln{\mathcal{P}_T}}{dN}=-2\epsilon_H+\frac{du_h}{dh}\frac{1+2h^2}{h}\left(\eta_H-\eta_L\right),
\end{equation}
where we have used $N\simeq \ln{k}$ and (\ref{hk}). We can see even when the energy of gauge fields is not so small $h>\mathcal{O}(1)$ the spectral index can be suppressed by small $du_h/dh\sim\mathcal{O}(hL)$ hence the power spectrum is almost scale invariant. The tensor-to-scalar ratio is given by
\begin{equation}
    r\equiv\frac{\mathcal{P}_T}{\mathcal{P}_{\mathcal{R}}}=32\epsilon_H\frac{h^2\left(1+2h^2\right)}{\left(3+2h^2\right)^2}e^{2u_h-2u_Q}.
\end{equation}
We have already known $u_Q$ is linear with respect to $ph$ for $h>\mathcal{O}(1)$ while the $u_h$ is of order of $hL\ll ph$. Hence, the tensor-to-scalar ratio $r$ is exponentially suppressed if $h\gg1$. This is similar to the hyperbolic inflation whose tensor-to-scalar ratio is suppressed by quick rotation of the angular field \cite{Mizuno:2017idt}. Although we have not yet detected the primordial gravitational waves, if we can see them in the near future one should not expect the viability of hyperbolic inflation and our model with large $h$. However, as we mentioned above, there also exist attractors for $h\sim\mathcal{O}(1)$, where $r$ need not be suppressed. But we also have a lower bound $h>0.3$, which guarantees the scale invariance of the CMB power spectrum.

\subsection{Two-form gauge fields}
We first calculate the curvature perturbations of the two-form case. In this case there is only one physical degree of freedom of scalar perturbation $\mathbb{B}$ for the gauge fields. We take the spatially flat gauge $\psi=0$. 
The curvature perturbations given by (\ref{R}), where $T^0_i=-\dot{\phi}\partial_i\delta\phi-(g^2\dot{\mathbb{B}}/a^4)\partial_i\delta\mathbb{B}$ and $\rho+p=\dot{\phi}^2+g^2\dot{\mathbb{B}}^2/a^4$ read
\begin{equation}
    \mathcal{R}=-H\frac{\dot{\phi}\delta\phi+(g^2\dot{\mathbb{B}}/a^4)\delta\mathbb{B}}{\dot{\phi}^2+g^2\dot{\mathbb{B}}^2/a^4}.
\end{equation}
For $l>0.6$ with only $d_1$ modes contributing to the $\mathcal{R}$, we have
\begin{equation}
    \mathcal{R}=-\frac{1}{aL}\frac{-\Delta_{\phi}+l\Delta_{P}}{1+l^2}=-\frac{1}{L_2}\frac{l^2-3}{l\left(1+l^2\right)}\frac{\Delta_{P}}{a},
\end{equation}
where we have use $\Delta_{\phi}/\Delta_{P}=3/l$. We find the perturbations $\Delta_{\phi}$ and $\Delta_{P}$ have the same sign, which will cancel to each other in the curvature $\mathcal{R}$ when $l=\sqrt{3}$. Similar to the one-form case, the amplitude of the perturbation is given by $\tilde{\tau}^2\mathcal{P}_{\Delta_{P}}(l=0)=1/(2\pi^2)$. Then after using (\ref{h}) and $\tilde{\tau}^2\mathcal{P}_{\tilde{\Delta}_{i}}=\tau^2\mathcal{P}_{\Delta_{i}}$ the power spectrum of curvature is given by
\begin{equation}
    \mathcal{P}_{\mathcal{R}}=\frac{H^2}{L_2^2}\left(\frac{l^2-3}{l\left(1+l^2\right)}\right)^2\tau^2 \mathcal{P}_{\Delta_P}=\frac{1}{(2\pi)^2}\frac{H^2}{2\epsilon_H}\frac{\left(l^2-3\right)^2}{l^2\left(1+l^2\right)}e^{2v_{P}(l)},
\end{equation}
where we have used $\epsilon_H=L_2^2/2+\epsilon_B/2=L_2^2(1+l^2)/2$ in the second equality. The power spectrum of $\mathcal{R}$ is also scale invariant on large scales ((see Figure \ref{fig:PowerSpectrum} ($Right$))). Also, $\mathcal{P}_{\mathcal{R}}$ has a minimum at $l=\sqrt{3}$. When the mode with momentum $k$ crosses the horizon, we have
\begin{equation}
    \frac{dl^2}{d\ln{k}}\simeq(1+l^2)(\eta_H-\eta_L).
\end{equation}
Then the spectral index is calculated as
\begin{equation}
    n_s-1\simeq -2\epsilon_H-\eta_H+\left(\frac{2+2l^2}{l^2-3}-\frac{1+2l^2}{l^2}+\frac{dv_{P}}{dl}\frac{1+l^2}{l}\right)\left(\eta_H-\eta_L\right).
\end{equation}
We see the first term will be diverge at $l=\sqrt{3}$. For large $l$, the first two term in the bracket will be suppressed so $n_s-1\simeq-2\epsilon_H-\eta_H+r_2l(\eta_H-\eta_L)$. 
There is also an exponential suppression in the ratio,
\begin{equation}
    r=16\epsilon_H\frac{l^2\left(1+l^2\right)}{\left(l^2-3\right)^2}e^{-2v_P}.
\end{equation}
Nevertheless, we have mentioned before that $v_P$ does not change significantly even when $l$ is very large ($v_{P}\sim\mathcal{O}(0.1)$ when $l\sim\mathcal{O}(10^3)$). That is, for $l\gg 1$, we still have
\begin{equation}
r\sim \mathcal{O}(1)\epsilon_H.
\end{equation}
Hence the tensor-to-scalar ratio would not be overly suppressed in the two-form case. This can be distinguished from the one-form case, and also the hyperbolic inflation. Moreover, the ratio will become large when $l\simeq\sqrt{3}$. For this value, the ratio and the spectral index become very large, which should be ruled out.

\section{Summary}
It is known that in some UV theory at high energy scales, the extra heavy scalar fields can contribute to the single-field inflation through the non-canonical kinetic terms of the multi-scalar fields. The effect of negative curvature of such field space will cause the instability to the entropic perturbation during inflation and finally make a transition from  the conventional slow-roll inflation to another attractor. In this paper, we have extended the geometric destabilization of multi-scalar-field inflation
 to multi-form field inflation. More precisely, we have explored the cases that the inflaton is coupled to extra massless gauge fields with  nontrivial kinetic terms of these fields (\ref{actionall}), which is originally motivated by supergravity. We discussed the coupling with isotropic  one-form and two-form gauge fields separately.

In section \ref{GD} we used the helicity decomposition and derived the equations of motion of scalar  perturbations of the gauge fields. We found these perturbations become tachyonic, i.e., the mass squared terms (\ref{MQ2}) and (\ref{MP2}) become negative if the effect of geometry of fields is large enough. This is the geometric destabilization of gauge fields. We provide some examples of such destabilization in section \ref{examples}, including the known anisotropic inflation, which only contains one scalar field \cite{Watanabe:2009ct} and the anisotropic hyperbolic inflation, which contains two scalar fields \cite{Chen:2021nkf}. Similar to the multi-scalar-field case, the system exhibits a transition to a second inflationary phase after the destabilization due to gauge fields. In section \ref{Aad}, we found that for many different choices of metric in the field space, the mass squared terms of perturbations always become almost zero after the back-reaction becomes important (Figure. \ref{fig:MU} and Figure. \ref{fig:4}). We found the zero mass squared term provides a general form of attractors of the second inflationary phase,
\begin{equation}
    \dot{\phi}=-HL(\phi).
\end{equation}
which is very similar to that of the hyperbolic inflation \cite{Brown:2017osf}. 
The quantity $L(\phi)$ is slowly varying and determined by the metric in field space. In other words, if the potential is steep enough, the conventional slow-roll attractor will be destabilized and make a transition to a new attractor. In this new inflationary phase, the evolution of inflaton $\dot{\phi}/H$ is given by $L(\phi)$ instead of the shape of potential. We also found in these attractors, the metric in field space becomes hyperbolic type (\ref{hyperbolic}). But in contrast to the hyperbolic inflation, the energy of extra fields (in our case are gauge fields) can be moderate compared to the kinetic energy of inflaton (i.e., $h,l\sim\mathcal{O}(1)$ are allowed) so that the tensor-to-scalar ratio is not significantly  suppressed.

We then explored the perturbation of these models in section \ref{perturbation} and section \ref{PPS}. Thanks to the slow roll, one can treat the energy ratio $h$ and $l$ as parameters of these models in a de Sitter background. We discussed the dynamics of the perturbations of one-form and two-form cases separately. We considered the regime that $h$ and $l$ are not so small, hence we directly solved the equations of motion without resorting to the in-in formalism. 
The main results of this part are as follows:
\begin{itemize}

\item[$\bullet$]
The dynamics of scalar perturbations were divided into three cases $h\ll1$ ,$h\lesssim0.3$ and $h>0.3$ for one-form and $l\ll1$, $l\lesssim0.6$ and $l>0.6$ for two-form respectively. For $h>0.3$ and $l>0.6$, the massive modes decay rapidly hence $\tau\Delta_{\phi}$, $\tau\Delta_{Q}$ and $\tau\Delta_{P}$ are constant on super-horizon scale. However, if $h\lesssim0.3$ and $l\lesssim0.6$, the massive modes decay slowly hence these scalar modes deviate from  constants. When $h,l\ll1$ the massive modes decay very slowly. Hence these scalar modes become nearly constant again.
\item[$\bullet$] 
As in the hyperbolic inflation, the scalar modes experience an exponential growth before  the horizon crossing. We numerically calculate the growth factor. For one-form case, the scalar mode $\Delta_Q$ exponentially grows with increasing $h$, i.e., $\sim e^{ph+q}$ when $h>\mathcal{O}(1)$. But for the two-form case, the scalar mode $\Delta_P$ barely grows with increasing $l$ when $l>\mathcal{O}(1)$ so the growth factor is about $e^{\mathcal{O}(10^{-1})}\sim \mathcal{O}(1)$ even for very large energy density of gauge fields. 
\item[$\bullet$] 
The primordial power spectrum of curvature perturbation with $h>0.3$ for one-form and $l>0.6$ for two-form case were also calculated. The growth of tensor modes in these two case are insignificant compared to the scalar one hence the modifications to power spectrum of tensor modes are negligible. Therefore for the one-form case, the tensor-to-scalar ratio $r$ will be exponentially suppressed by this enhancement. While for the two-form case, the suppression is insignificant because the growth factor of $\mathcal{R}$ stops at $e^{\mathcal{O}(10^{-1})}$. We also found that in contrast to the one-form case, the fluctuations of $\Delta_\phi$ and $\Delta_{P}$ in two-form case could cancel to each other in curvature perturbation, i.e., $\mathcal{R}=0$ when $l=\sqrt{3}$. It will 
lead to divergences in the spectral index $n_s$ and ratio $r$, which should be ruled out.
\end{itemize}

In this paper we just got a preliminary glimpse of the geometric destabilization of gauge fields. Although we considered isotropic configurations of gauge fields, which can be treated as a scalar field in the background evolution. The situation is more complicate than the multi-scalar-field case. The field space of our case includes scalar fields and gauge fields at the same time. But the evolution of these two kind of fields are very different (scalar fields obey Klein-Gordon equations while gauge fields obey Maxwell equations at classical level). So unlike the multi-scalar-field case, we have not written down the covariant form of the perturbation equations of motion. We directly study the mass squared terms of perturbed equations of gauge fields, not the entropic one. Moreover, we also found some differences between one-form and two-form cases at the perturbation level. So it is necessary to find a more general description of these models. 

Although we found some general features of the attractors after geometric destabilization, the details of this transition are still unclear because the perturbative methods is invalid for the large back-reaction from perturbations. It is also interesting to study a full non-linear effective theory of inflation. One thing worth mentioning is that we had no a priori restrictions on the metric of the field space at the beginning. Nevertheless, the field space will become hyperbolic plane (or exponential type) at leading order of slow roll approximation. We have known that the hyperbolic geometry of multi-scalar-field is quite interesting because of its maximal symmetries and relation with $\alpha$-attractor model. Whether hyperbolic plane plays an important role in these models can be explored in the future.

In this paper, we calculate the primordial perturbations and power spectrum in the presence of one-form and two-form cases. It is also interesting to study the primordial bispectrum then the non-Gaussianities of these models. 
As we mentioned in the introduction, the non-Gaussianities of hyperbolic inflation is quite different from that of the single-field inflation. We can expect that the attractors in the multi-form-field inflation enjoy similar features to the hyperbolic inflation. We leave all of these questions for future studies.

\acknowledgments

We would like to thank Kimihiro Nomura and Pak Hang Chris Lau for helpful discussions. J.\,S. was in part supported by JSPS KAKENHI Grant Numbers JP17H02894, JP17K18778, JP20H01902. C-B.\,C. was supported by Japanese Government (MEXT) Scholarship and China Scholarship Council (CSC).

\appendix
\section{Quadratic action}
In this appendix, we provide details of derivation of the quadratic action and equations of motion of perturbations in our paper. 

Let us consider the Arnowitt-Deser-Misner(ADM) form of the metric
\begin{equation}\label{bgv1}
    ds^2=-\mathcal{N}^2dt^2+q_{ij}\left(N^idt+dx^i\right)\left(N^jdt+dx^j\right),
\end{equation}
where $\mathcal{N}$ is the lapse function, $N^i$ is the shift vector and $q_{ij}$ is the metric of hypersurface. Note that, for background quantities, $\mathcal{N}=1$, $N^i=0$ and $q_{ij}=a^2(t)\delta_{ij}$.
Using ADM formalism, we can write down the action of gravity and scalar field as the following form
\begin{align}
    S_{\text{\text{gravity}}}&=\int d^4xN\sqrt{q}\left[\frac{M_{\text{pl}}^2}{2}\left(^{(3)}R+K_{ij}K^{ij}-K^2\right)\right],\\
    S_{\text{scalar}}&=\int d^4x\sqrt{q}\left[\frac{1}{2N}\pi^2-\frac{N}{2}\partial_i\phi\partial^i\phi-NV(\phi)\right],
\end{align}
where we have defined $\pi=\dot{\phi}-N^j\phi_{|j}$ and $q_{ij}$ is the induced metric of space-like hypersurface. The extrinsic curvature of this hypersurface is definied by 
\begin{equation}
K_{ij}=\frac{1}{2N}\left(\dot{q}_{ij}-2N_{(i|j)}\right)
\end{equation}
and the Ricci scalar of this hypersurface is given by
\begin{equation}
    {^{(3)}R}=\left(q_{ij,kl}+q_{mn}{^{(3)}\Gamma^m_{ij}} {^{(3)}\Gamma^n_{kl}}\right)\left(q^{ik}q^{jl}-q^{ij}q^{kl}\right),
\end{equation}
where ${^{(3)}\Gamma^i_{jk}}=q^{il}\left(q_{lj,k}+q_{lk,j}-q_{jk,l}\right)/2$ is the Christoffel symbol of the induced metric. 

\subsection{One-form gauge fields}\label{quadratic_action}
For the one-form gauge fields, the ADM formalism of the action is given by
\begin{align}
    S_{\text{gauge}}&=\int d^4x\sqrt{q}\left[\frac{f_{ab}}{2N}q^{ik}(E^a_{\ i}+F^a_{\ ij}N^j)(E^b_{\ k}+F^b_{\ kl}N^l)-\frac{N}{4}f_{ab}q^{ik}q^{jl}F^a_{ij}F^b_{kl}\right],
\end{align}
where $E^a_{\ i}\equiv F^a_{\ 0i}$. Without loss of generality, here we consider a model with only one scalar field $\phi$ and a triplet of gauge fields ($A^a_{\ i},\ a=1,2,3$), which allows an isotropic background of the theory. This can be realized by three orthogonal $U(1)$ vector fields 
\begin{equation}\label{bgv2a}
A^a_{\ 0}=0,\ \ \ \ \ A^a_{\ i}=\mathbb{A}\delta_{ai}, \ \ \ \ \ f_{ab}=f^2\delta_{ab},
\end{equation}
which enjoys internal global $O(3)$ symmetry of the gauge-field space. In other words, the rotational can be absorbed by the internal $O(3)$ transformation of gauge fields, which admits isotropic FLRW background. It suggests that one can use the  decomposition of both metric and gauge fields into the scalar, vector and tensor modes under $O(3)$. There are two physical degrees of freedom in one one-form gauge field so we should have six physical degrees of freedom in our configuration of gauge fields.

We perturb the quantities (\ref{bgv1}) and (\ref{bgv2a}) as follow. First, the metric and scalar fields are decomposed as
\begin{equation}
\mathcal{N}=1+A,\ \ \ \ N_i=\partial_iB+B_i,\ \ \ \ q_{ij}=a^2(t)(\delta_{ij}+\gamma_{ij}),\ \ \ \ \phi=\phi(t)+\delta\phi
\end{equation}
where
\begin{equation}
    \gamma_{ij}=-2\psi\delta_{ij}+2E_{,ij}+2W_{(i,j)}+w_{ij}.
\end{equation}
Here $B_i$ and $W_i$ are transverse vector ($\partial^iB_i=\partial^iW_i=0$) and $t_{ij}$ is a transverse traceless tensor ($\partial^it_{ij}=t^i_{\ i}=0$). We also decompose the gauge fields as follow
\begin{align}\label{Gdecompose}
A^a_{\ 0}&=\partial_a\mathbb{Y}+\mathbb{Y}_a,\nonumber\\
A^a_{\ i}&=\left(\mathbb{A}+\delta\mathbb{A}\right)\delta_{ai}+\epsilon_{iab}\left(\partial_b\mathbb{U}+\mathbb{U}_b\right)+\partial_i\partial_a\mathbb{M}+\partial_{(i}\mathbb{M}_{a)}+\mathbb{T}_{ai},
\end{align}
where $\mathbb{Y}_a$, $\mathbb{U}_a$ and $\mathbb{M}_a$ are all transverse vector ($\partial^a\mathbb{Y}_a=\partial^a\mathbb{U}_a=\partial^a\mathbb{M}_a=0$) and $\mathbb{T}_{ai}$ is a symmetric transverse traceless tensor ($\partial{^a}\mathbb{T}_{ai}=\mathbb{T}^a_{\ a}=0$). We do not distinguish indices $a$ and $i$ here because we identified the spatial rotation symmetry with the internal global $O(3)$ symmetry of the space of  the gauge fields. 

Now we have scalar modes ($A,B,\psi,E,\delta\phi,\mathbb{Y},\delta\mathbb{A},\mathbb{U},\mathbb{M}$), four from metric, one from scalar field and four from gauge fields. The vector modes are ($B_i,W_i,\mathbb{Y}_a,\mathbb{U}_a,\mathbb{M}_a)$, two from metric and three from gauge fields. The tensor modes are ($t_{ij},T_{ai}$), one from metric and another from gauge fields. However, not all these quantities are physical because there exist gauge redundancy in metric and gauge fields. First we choose the spatially flat gauge 
\begin{equation}
    \psi=0,\ \ \ \ \ E=0 \ \ \ \ \ \text{and} \ \ \ \ \ W_i=0
\end{equation}
to fix the metric modes. In the gauge fields, we will see $\mathbb{Y}$ and $\mathbb{Y}_a$ are non-dynamical hence can be eliminated. We have only nine$(=3+2\times 2+2)$ dynamical modes $(\delta\mathbb{A},\mathbb{U},\mathbb{M},\mathbb{U}_b,\mathbb{M}_a,\mathbb{T}_{ai})$. The gauge fields enjoys local $U(1)$ symmetry 
\begin{align}
    A^a_{\ \mu}\rightarrow A^a_{\ \mu}+\partial_{\mu}\rho^a,
\end{align}
where $\rho^a$ is a arbitrary function. It can be also decomposed as 
\begin{equation}
    \rho^a=\partial_a\rho+\rho_a
\end{equation}
with $\partial^a\rho_a=0$. This yields the local symmetry of perturbations
$\delta A^a_{\ \mu}\rightarrow \delta A^a_{\ \mu}+\partial_{\mu}\partial_a \rho+\partial_{\mu}\rho_a$. 
From (\ref{Gdecompose}) we find the scalar and vector modes may transform as
\begin{align}
    \text{Scalar:}\ \ \ &\mathbb{Y}\rightarrow \mathbb{Y}+\dot{\rho}, \ \ \ \ \ \mathbb{M}\rightarrow \mathbb{M}+\rho,\\
    \text{Vector:}\ \ \ &\mathbb{U}_b\rightarrow \mathbb{U}_b+\frac{1}{2}\epsilon_{iab}\partial_i\rho_a
\end{align}
while the other modes are invariant under transformation. So there are still three redundancy and we can fix them by
\begin{equation}
    \mathbb{M}=0\ \ \ \ \text{and} \ \ \ \ \mathbb{U}_b=0.
\end{equation}
Now the rest perturbations of gauge fields are ($\delta\mathbb{A},\mathbb{U},\mathbb{M}_a,\mathbb{T}_{ai}$). There are six dynamical degrees of freedom we need to consider in the one-form gauge fields.

Expanding the action with one scalar field around background (\ref{bgv1}) and (\ref{bgv2}) up to second order of scalar modes only, and then using background equations of motion and performing several integration by parts yields the following quadratic action of perturbations
\begin{align}\label{action2}
    S^{(2)}_{\text{scalar}}=\frac{1}{2}&\int dtd^3xa^3\Bigg\{\Big[-6M_{\text{pl}}^2H^2A+\dot{\phi}^2A-2V_{\phi}\delta\phi-2\dot{\phi}\delta\dot{\phi}+3f^2\frac{\dot{\mathbb{A}}^2}{a^2}A-6ff_{\phi}\frac{\dot{\mathbb{A}}^2}{a^2}\delta\phi\nonumber\\
    &-2f^2\frac{\dot{\mathbb{A}}}{a^2}\big(3\delta\dot{ \mathbb{A}}-\partial^2\mathbb{Y}\big) \Big]A+\frac{2}{a^2}\Big(\dot{\phi}\delta\phi-2M_{\text{pl}}^2HA+2f^2\frac{\dot{\mathbb{A}}}{a^2}\delta \mathbb{A}\Big)\partial^2B\nonumber\\
    &+(\delta\dot{\phi})^2-\frac{1}{a^2}(\partial_i\delta\phi)^2-V_{\phi\phi}\left(\delta\phi\right)^2+\frac{f^2}{a^2}\Big[3(\delta\dot{ \mathbb{A}})^2-\frac{2}{a^2}(\partial_i \delta \mathbb{A})^2\nonumber\\
    &+2(\partial_i\dot{\mathbb{U}})^2-\frac{2}{a^2}\partial^2\mathbb{U}\partial^2\mathbb{U}+\partial^2\mathbb{Y}\big(\partial^2\mathbb{Y}-2\delta\dot{ \mathbb{A}}\big)\Big]\nonumber\\
    &+\frac{4}{a^2}ff_{\phi}\dot{\mathbb{A}}\big(3\delta\dot{\mathbb{A}}-\partial^2\mathbb{Y}\big)\delta\phi+\frac{3}{a^2}\big(f_{\phi}^2+ff_{\phi\phi}\big)\mathbb{A}^2(\delta\phi)^2\Bigg\},
\end{align}
where we have denoted $\partial^2\equiv\partial_i\partial_i$. Although we consider isotropic variables of gauge fields, like scalar fields, what different from the case that contains two scalar fields is that, there are also interactions with perturbations of the temporal gauge $\mathbb{Y}$. It results different dynamics of these perturbations.

We have already mentioned that modes $A$, $B$ and $\mathbb{Y}$ appear with no time derivative which means that they are non-dynamical degrees of freedom and can be eliminated. Varying the quadratic action with respect to $A$, $B$ and $\mathbb{Y}$ respectively yield
\begin{align}
    \dot{\phi}\delta\phi=&2M_{\text{pl}}^2HA-2f^2\frac{\dot{\mathbb{A}}}{a^2}\delta\mathbb{A},\\
    -2M_{\text{pl}}^2\frac{H}{a^2}\partial^2B=&6M_{\text{pl}}^2H^2A+\dot{\phi}\delta\dot{\phi}+V_{\phi}\delta\phi-\dot{\phi}^2A-3f^2\frac{\dot{\mathbb{A}}^2}{a^2}A\nonumber\\
    &+f^2\frac{\dot{\mathbb{A}}}{a^2}\big(3\delta\dot{\mathbb{A}}-\partial^2Y\big)+3ff_{\phi}\frac{\dot{\mathbb{A}}}{a^2}\delta\phi,\\
    \partial^2\mathbb{Y}=&\delta\dot{\mathbb{A}}+2\frac{f_{\phi}}{f}\dot{\mathbb{A}}\delta\phi-\dot{\mathbb{A}}A.
\end{align}
After substituting these equations into (\ref{action2}) we can obtain the quadratic action for the remaining dynamical modes. Before doing that, we introduce new variable of the gauge fields
\begin{equation}
\mathbb{A}\equiv a(t)\left(\tilde{Q}+\delta\tilde{Q}\right),\ \ \ \ \ \mathbb{U}\equiv a(t)\tilde{U},
\end{equation}
because the new quantity $\tilde{Q}$ and $\tilde{U}$ transform like scalar fields under rotation. Moreover, to obtain the equation of motion of gauge-field modes that looks like one of the scalar-field modes $\delta\phi$, we can introduce new variables of perturbations of gauge fields as
\begin{equation}
    \delta Q\equiv\sqrt{2}f\delta \tilde{Q},\ \ \ \ \ \ U_i\equiv\sqrt{2}f \partial_i\tilde{U}.
\end{equation}
We will see how $\delta Q$ and $\delta U$ destabilize the conventional slow-roll inflation.

After using these variables, background equations of motion and performing several integration by parts, we derive the final quadratic action
\begin{align}\label{action2f}
    S^{(2)}_{\text{scalar}}=\frac{1}{2}&\int dtd^3xa^3\Bigg\{\delta\dot{\phi}^2-\frac{1}{a^2}(\partial_i\delta\phi)^2+\Big[-V_{\phi\phi}+2H^2\big(3+\epsilon_H\big)\epsilon+4H^2\eta\epsilon\nonumber\\
    &+M_{\text{pl}}^2H^2\epsilon_E\Big(3\frac{f_{\phi\phi}}{f}-\frac{f_{\phi}^2}{f^2}-4\sqrt{2}\frac{\sqrt{\epsilon}}{M_{\text{pl}}}\frac{f_{\phi}}{f}\Big)\Big](\delta\phi)^2+(\delta \dot{Q})^2-\frac{1}{a^2}(\partial_i\delta Q)^2\nonumber\\
    &+\Big[\frac{\ddot{f}}{f}+\frac{H\dot{f}}{f}-H^2\big(2-\epsilon_H\big)-2H^2(3-\epsilon_H)\epsilon_E\Big](\delta Q)^2+4H^2\sqrt{\epsilon_E}\Big[\sqrt{\epsilon}\big(\epsilon_H+\eta\big)\nonumber\\
    &+\sqrt{2}M_{\text{pl}}\frac{f_{\phi}}{f}\big(1-\epsilon_E-\frac{\dot{f}}{Hf}\big)\Big]\delta\phi\delta Q+4\sqrt{2}M_{\text{pl}}H\frac{f_{\phi}}{f}\sqrt{\epsilon_E}\delta\dot{ Q}\delta\phi+(\dot{U}_i)^2\nonumber\\
    &-\frac{1}{a^2}(\partial_j U_i)^2
    +\Big[\frac{\ddot{f}}{f}+\frac{H\dot{f}}{f}-H^2\big(2-\epsilon_H\big)\Big](U_i)^2\Bigg\},
\end{align}
where we have defined dimensionless parameters $\sqrt{\epsilon}\equiv\dot{\phi}/(\sqrt{2}M_{\text{pl}}H)$, $\sqrt{\epsilon_E}\equiv f\dot{\mathbb{A}}/(M_{\text{pl}}aH)$ and $\epsilon_H\equiv -\dot{H}/H^2$. We mention that here the root is just a symbol and doesn't mean that $\sqrt{\epsilon}$ is positive. Varying the quadratic action with respect to $\delta Q$ and $U$ yields their equations of motion
\begin{align}
    &\delta \ddot{Q}+3H\delta \dot{Q}+\left(M_{QQ}+\frac{k^2}{a^2}\right)\delta Q+M_{Q\phi}\delta\phi+\bar{M}_{Q\phi}\dot{\delta\phi}=0,\\
    &\ddot{ U}_i+3H\dot{ U}_i+\left(M_{UU}+\frac{k^2}{a^2}\right) U_i=0,
\end{align}
where we have defined 
\begin{align}
    M_{QQ}\equiv& 2\epsilon_E\left(3-\epsilon_H\right)H^2-\left(\frac{\ddot{f}}{f}+H\frac{\dot{f}}{f}-2H^2+\epsilon_HH^2\right),\label{MQ}\\
    M_{UU}\equiv& -\left(\frac{\ddot{f}}{f}+H\frac{\dot{f}}{f}-2H^2+\epsilon_HH^2\right),\\
    M_{Q\phi}\equiv& 2\sqrt{\epsilon_E}\left[\sqrt{2}\epsilon_EM_{\text{pl}}\frac{f_{\phi}}{f}-2\sqrt{\epsilon}\left(\epsilon_H+\eta\right)
    +\frac{\sqrt{2}M_{\text{pl}}}{H}\left(\frac{\dot{f}_{\phi}}{f}-\frac{f_{\phi}\dot{f}}{f^2}\right)\right]H^2,\\
    \bar{M}_{Q\phi}\equiv& 2\sqrt{2}\sqrt{\epsilon_E}M_{\text{pl}}H\frac{f_{\phi}}{f}.
\end{align}
When energy density of gauge fields is negligible, i.e., $\epsilon_E=0$, the interactions between scalar-field modes and gauge-field modes $M_{Q\phi}=\bar{M}_{Q\phi}=0$. In addition, if we consider the slow roll inflation and small enough energy density of gauge fields, the first term in (\ref{MQ}) can be ignored compared to the second term. Hence perturbations $\delta Q$ and $U$ have the same effective mass squared $M_{QQ}\simeq M_{UU}$.

Next we are going to treat the vector sector. There are two vector perturbations $(\mathbb{Y}_a,\mathbb{M}_a)$ in the one-form gauge fields and one $B_j$ in the gravity after gauge-fixing. Expanding the action around background and write down the quadratic action by performing several integration by parts we have
\begin{align}\label{vaction2}
    S^{(2)}_{\text{vector}}=\int dtd^3x&\Bigg\{\frac{M_{\text{pl}}^2}{4a}\partial_iB_j\partial_iB_j+\frac{af^2}{2}\Big(2\partial_i\dot{\mathbb{M}}_a\partial_i\dot{\mathbb{M}}_a-2\partial_i\dot{\mathbb{M}}_a\partial_i\mathbb{Y}_a+\partial_i\mathbb{Y}_a\partial_i\mathbb{Y}_a\nonumber\\
    &-\frac{2\dot{\mathbb{A}}}{a^2}\partial_i\mathbb{M}_j\partial_iB_j\Big)-\frac{f^2}{2a}\partial_k(\partial_i\mathbb{M}_a)\partial_k(\partial_i\mathbb{M}_a)
    \Bigg\}.
\end{align}
We can see $B_j$ and $\mathbb{Y}_a$ are non-dynamical hence we can obtain their equations of motion by varying the quadratic action with respect to these two perturbations
\begin{align}
    \mathbb{Y}_a&=\dot{\mathbb{M}}_a,\\
    B_j&=\frac{2f^2\dot{\mathbb{A}}}{M_{\text{pl}}^2}\mathbb{M}_j.
\end{align}
After substituting these equations into quadratic action of vector sector (\ref{vaction2}) we finally have the action of dynamical perturbations $M_a$
\begin{align}\label{action2v}
    S^{(2)}_{\text{vector}}=\int dtd^3x&\Bigg\{\frac{af^2}{2}\partial_i\dot{\mathbb{M}}_a\partial_i\dot{\mathbb{M}}_a-\frac{f^2}{2a}\partial_k(\partial_i\mathbb{M}_a)\partial_k(\partial_i\mathbb{M}_a)-\frac{f^4\dot{\mathbb{A}}^2}{M_{\text{pl}}^2a}\partial_i\mathbb{M}_a\partial_i\mathbb{M}_a
\Bigg\}.
\end{align}

Finally we consider the tensor sector of the models. There are two dynamical perturbations $(w_{ij},\mathbb{T}_{ij})$ and no non-dynamical perturbation in the tensor part of the system. So we can directly expand the action around background and write down the quadratic action by performing several integration by parts
\begin{align}
    S^{(2)}_{\text{tensor}}=\int dt&d^3xa^3\Bigg\{\frac{M_{\text{pl}}^2}{8}(\dot{w}_{ij})^2-\frac{M_{\text{pl}}^2}{8a^2}(\partial_kw_{ij})^2+\frac{f^2}{2a^2}(\dot{\mathbb{T}}_{ij})^2-\frac{f^2}{2a^4}(\partial_k\mathbb{T}_{ij})^2\nonumber\\
    &+\frac{M_{\text{pl}}^2}{4}\frac{\dot{a}}{a}w_{ij}\dot{w_{ij}}-\frac{f^2\dot{\mathbb{A}}}{a^2} w_{ij}\dot{\mathbb{T}}_{ij}+\Big(\frac{M_{\text{pl}}^2}{8}\frac{\ddot{a}}{a}+\frac{M_{\text{pl}}^2}{4}\frac{\dot{a}^2}{a^2}+\frac{\dot{\mathbb{A}}^2}{4a^2}f^2\Big)(w_{ij})^2\Bigg\}.
\end{align}
In order to leave quadratic action with only background gauge fields we also have substituted the background equations of motion to eliminate $\dot{\phi}^2$ and $V(\phi)$ above.

\subsection{Two-form gauge fields}\label{2quadratic_action}
For the Two-form gauge fields, the ADM formalism of the action is given by
\begin{align}
    S_{\text{gauge}}=\int d^4x\sqrt{q}&\bigg[\frac{1}{4N}g_{ab}q^{ik}q^{jl}(I^a_{\ ij}-H^a_{\ ijm}N^m)(I^b_{\ kl}-H^b_{\ kln}N^n)\nonumber\\
    &-\frac{N}{12}g_{ab}q^{mn}q^{ik}q^{jl}H^a_{\ ijm}H^b_{\ kln}\bigg],
\end{align}
where $I^a_{\ ij}\equiv H^a_{\ 0ij}$ and the field strength $H^a_{\ \mu\nu\lambda}\equiv \partial_{\mu}A^a_{\ \nu\lambda}+\partial_{\nu}A^a_{\ \lambda\mu}+\partial_{\lambda}A^a_{\ \mu\nu}$. We can spatially dualize the two-form gauge fields $A^a_{\ ij}$ as follow \cite{Germani:2009iq}
\begin{equation}
    A^a_{\ ij}\equiv \epsilon_{ijk}B^a_{\ k}.
\end{equation}
Similarly to the one-form gauge fields, we also consider an isotropic triplet of the duality of the background tow-form gauge fields ($B^a_{\ i}$, $a=1,2,3$), which is realized by three orthogonal two-form fields
\begin{equation}\label{bg2v2a}
    A^a_{\ 0i}=0,\ \ \ \ \ \ B^a_{\ k}=\mathbb{B}\delta_{ai},\ \ \ \ \ \ g_{ab}=g^2\delta_{ab},
\end{equation}
where we have fixed the gauge of $B^a_{\ 0i}=0$ for background configuration and $B^a_{\ k}$ enjoy the rotational symmetry. A two-form gauge field can be represented by a scalar field hence we should have three physical degrees of freedom in this configuration. One can write the two-form gauge fields as
\begin{align}\label{G2decompose}
    A^a_{\ 0i}&=\mathbb{D}\delta_{ai}+\epsilon_{iab}\left(\partial_b\mathbb{X}+\mathbb{X}_b\right)+\partial_i\partial_a\mathbb{Z}+\partial_{(i}\mathbb{Z}_{a)}+\mathbb{R}_{ai},\nonumber\\
    B^a_{\ k}&=\left(\mathbb{B}+\delta\mathbb{B}\right)\delta_{ak}+\epsilon_{kab}\left(\partial_b\mathbb{V}+\mathbb{V}_b\right)+\partial_{k}\partial_{a}\mathbb{W}+\partial_{(k}\mathbb{W}_{a)}+\mathbb{S}_{ak},
\end{align}
where $\mathbb{X}_a$, $\mathbb{Z}_a$, $\mathbb{V}_a$ and $\mathbb{W}_a$ are all transeverse vector ($\partial^a\mathbb{X}_a=\partial^a\mathbb{Z}_a=\partial^a\mathbb{V}_a=\partial^a\mathbb{W}_a=0$) and $\mathbb{R}_{ai}$ and $\mathbb{S}_{ai}$ are symmetric transeverse traceless tensors ($\partial^a\mathbb{R}_{ai}=\mathbb{R}^a_{\ a}=\partial^a\mathbb{S}_{ai}=\mathbb{S}^a_{\ a}=0$). Also, we do not distinguish indices because we identified spatial indices and internal space indices.

Let us fix the gauge ambiguity of the theory. We choose the spatially flat gauge $\psi=E=W_i=0$. For the two-form gauge fields, nine$(=3+2\times 2+2)$ modes $(\mathbb{D},\mathbb{X},\mathbb{Z},\mathbb{X}_b,\mathbb{Z}_a,\mathbb{R}_{ai})$ are non-dynamical. We have nine$(=3+2\times 2+2)$ dynamical modes $(\delta\mathbb{B},\mathbb{V},\mathbb{W},\mathbb{V}_b,\mathbb{W}_a,\mathbb{S}_{ak})$. However, we also have to consider the gauge transformation 
$A^a_{\ \mu\nu}\rightarrow A^a_{\ \mu\nu}+2\partial_{\mu}\xi^a_{\ \nu}-2\partial_{\nu}\xi^a_{\ \mu}$ 
of two-form gauge fields. Equivalently, the gauge transformation of the spatial dual fields $B^a_{\ k}$ is given by 
\begin{equation}
    B^a_{\ k}\rightarrow B^a_{\ k}+\epsilon_{ijk}(\partial_{i}\xi^a_{\ j}-\partial_{j}\xi^a_{\ i}).
\end{equation}
We can also decompose the functions $\xi^a_{\ \mu}$ as
\begin{align}
    \xi^a_{\ 0}&=\partial_{a}\eta+\eta_a,\nonumber\\
    \xi^a_{\ j}&=\xi\delta_{aj}+\epsilon_{jab}\left(\partial_b\theta+\theta_b\right)+\partial_{j}\partial_{a}\zeta+\partial_{(j}\zeta_{a)}+\xi_{aj},
\end{align}
where $\eta_a$, $\theta_a$ and $\zeta_a$ are transverse vector ($\partial^a\eta_a=\partial^a\theta_a=\partial^a\zeta_a=0$) and $\xi_{ai}$ is symmetric transverse traceless tensor ($\partial^a\xi_{ai}=\xi^a_{\ a}=0$). From (\ref{G2decompose}) we can find the scalar, vector and tensor modes can transform as
\begin{align}
    \text{Scalar:}\ \ \ &\mathbb{D}\rightarrow\mathbb{D}+\dot{\xi},\ \ \ \ \  \mathbb{X}\rightarrow\mathbb{X}+\dot{\theta},\ \ \ \ \ \mathbb{Z}\rightarrow \mathbb{Z}+\dot{\xi}-\eta,\nonumber\\
    &\mathbb{V}\rightarrow\mathbb{V}-2\xi,\ \ \ \ \ \mathbb{W}\rightarrow\mathbb{W}+4\theta,\\
    \text{Vector:}\ \ \ &\mathbb{X}_b\rightarrow\mathbb{X}_b+\dot{\theta}_b-\frac{1}{2}\epsilon_{iab}\partial_i\eta_a,\ \ \ \ \ \mathbb{V}_b\rightarrow\mathbb{V}_b-\epsilon_{kab}\partial_a\theta_k+\frac{1}{2}\partial^2\zeta_b,\nonumber\\
    &\mathbb{Z}_a\rightarrow\mathbb{Z}_a+\dot{\zeta}_a,\\
    \text{Tensor:}\ \ \ &\mathbb{R}_{ai}\rightarrow\mathbb{R}_{ai}+\dot{\xi}_{ai}, \ \ \ \ \ \mathbb{S}_{ak}\rightarrow\mathbb{S}_{ak}+2\epsilon_{ijk}\partial_i\xi_{aj}.
\end{align}
while the other modes are invariant under transformation. We can see there are three redundancy for scalar modes, six redundancy for vector modes and two redundancy for tensor modes. So we can fix the gauge as
\begin{align}
    &\mathbb{W}=0,\ \ \ \ \ \ \mathbb{Z}=0,\ \ \ \ \ \ \mathbb{V}=0, \nonumber\\
    &\mathbb{Z}_a=0,\ \ \ \ \ \ \mathbb{V}_b=0,\ \ \ \ \ \ \mathbb{X}_b=0, \nonumber\\
    &\mathbb{S}_{ak}=0.
\end{align}
Moreover, the non-dynamical modes  $(\mathbb{D},\mathbb{X},\mathbb{X}_b,\mathbb{R}_{ai})$
can be eliminated from the quadratic action. Thus, there are only three dynamical degrees of freedom $(\delta\mathbb{B},\mathbb{W}_a)$ in the two-form gauge fields.

Expanding the action with one scalar field around background (\ref{bgv1}) and (\ref{bg2v2a}) up to the second order of scalar modes only, and then using background equations of motion and performing several integration by parts,
we obtain the following quadratic action 
\begin{align}\label{2action2}
    S^{(2)}_{\text{scalar}}=\frac{1}{2}&\int dtd^3xa^3\Bigg\{\Big[-6M_{\text{pl}}^2H^2A+\dot{\phi}^2A-2V_{\phi}\delta\phi-2\dot{\phi}\delta\dot{\phi}+3g^2\frac{\dot{\mathbb{B}}^2}{a^4}A-6gg_{\phi}\frac{\dot{\mathbb{B}}^2}{a^4}\delta\phi\nonumber\\
    &-2g^2\frac{\dot{\mathbb{B}}}{a^4}\big(3\delta\dot{\mathbb{B}}-2\partial^2\mathbb{X}\big)\Big]A+\frac{2}{a^2}\Big(\dot{\phi}\delta\phi-2M_{\text{pl}}^2HA+g^2\frac{\dot{\mathbb{B}}}{a^4}\delta\mathbb{B}\Big)\partial^2B+(\delta\dot{\phi})^2\nonumber\\
    &-\frac{1}{a^2}(\partial_i\delta\phi)^2-V_{\phi\phi}\left(\delta\phi\right)^2+\frac{g^2}{a^4}\Big[3(\delta\dot{\mathbb{B}})^2-\frac{1}{a^2}(\partial_i\delta\mathbb{B})^2+2(\partial_i\mathbb{D})^2-4\delta\dot{\mathbb{B}}\partial^2\mathbb{X}\nonumber\\
    &+2(\partial^2\mathbb{X})^2\Big]+\frac{4}{a^4}gg_{\phi}\dot{\mathbb{B}}(3\delta\dot{\mathbb{B}}-2\partial^2\mathbb{X})\delta\phi+\frac{3}{a^4}(g_{\phi}^2+gg_{\phi\phi})\dot{\mathbb{B}}^2(\delta\phi)^2
    \Bigg\}.
\end{align}
We note that modes $A$, $B$, $\mathbb{D}$ and $\mathbb{X}$ appear with no time derivative hence are non-dynamical degrees of freedom. After varying this quadratic action with respect to $A$, $B$, $\mathbb{D}$ and $\mathbb{X}$ respectively we obtain the constraint equations
\begin{align}
    \dot{\phi}\delta\phi=&2M_{\text{pl}}^2HA-\frac{g^2}{a^4}\dot{\mathbb{B}}\partial^2\mathbb{W},\\
    -2M_{\text{pl}}^2\frac{H}{a^2}\partial^2B=&6M_{\text{pl}}^2H^2A+\dot{\phi}\delta\dot{\phi}+V_{\phi}\delta\phi-\dot{\phi}^2A-3g^2\frac{\dot{\mathbb{B}}^2}{a^4}A\nonumber\\
    &+g^2\frac{\dot{\mathbb{B}}}{a^4}\big(\delta\dot{\mathbb{B}}-2\partial^2\mathbb{X}\big)+3gg_{\phi}\frac{\dot{\mathbb{B}}}{a^4}\delta\phi,\\
    \partial^2\mathbb{X}=&\delta\dot{\mathbb{B}}+2\frac{g_{\phi}}{h}\dot{\mathbb{B}}\delta\phi-\dot{\mathbb{B}}A,\\
    \mathbb{D}=&0.
\end{align}
The mode $\mathbb{D}$ vanishes. Similar to the one-form isotropic gauge fields,  there are interactions with perturbations of the temporal mode $\mathbb{X}$. To see the effective mass squared term of the equations of motion of gauge fields perturbations, one can define the new variables of the tow-form gauge fields as
\begin{equation}
    \delta P\equiv\frac{g}{a^2}\delta\mathbb{B}.
\end{equation}
After substituting constraint equations into (\ref{2action2}) and using the redefined variable $\delta P$, then performing several integration by parts and using background equations of motion, we obtain the quadratic action
\begin{align}
    S^{(2)}_{\text{scalar}}=\frac{1}{2}&\int dtd^3xa^3\Bigg\{\delta\dot{\phi}^2-\frac{1}{a^2}(\partial_i\delta\phi)^2+\Big[-V_{\phi\phi}+2H^2\big(3+\epsilon_H\big)\epsilon+4H^2\eta\epsilon\nonumber\\
    &+M_{\text{pl}}^2H^2\epsilon_B\Big(3\frac{g_{\phi\phi}}{g}-5\frac{g_{\phi}^2}{g^2}-2\sqrt{2}\frac{\sqrt{\epsilon}}{M_{\text{pl}}}\frac{g_{\phi}}{g}\Big)\Big](\delta\phi)^2+(\delta\dot{ P})^2-\frac{1}{a^2}(\partial_i\delta P)^2\nonumber\\
    &+\Big[\frac{\ddot{g}}{g}-\frac{H\dot{g}}{g}-2H^2(1-\epsilon_H)-(3-\epsilon_H)H^2\epsilon_B\Big](\delta P)^2+\Big[-2M_{\text{pl}}H^2\frac{g_{\phi}}{g}\sqrt{\epsilon_B}\nonumber\\
    &+2\sqrt{2}H^2\sqrt{\epsilon_B}\sqrt{\epsilon}(\epsilon_H+\eta)+\Big(2H-\frac{\dot{g}}{g}\Big)\Big(4M_{\text{pl}}H\frac{g_{\phi}}{g}\sqrt{\epsilon_B}\Big)\Big]\delta P\delta\phi\nonumber\\
    &+\Big(4M_{\text{pl}}H\frac{g_{\phi}}{g}\sqrt{\epsilon_B}\Big)\delta \dot{P}\delta\phi
    \Bigg\},
\end{align}
where we have defined dimensionless parameters $\sqrt{\epsilon}\equiv\dot{\phi}/(\sqrt{2}M_{\text{pl}}H)$, $\sqrt{\epsilon_B}\equiv g\dot{\mathbb{B}}/(M_{\text{pl}}a^2H)$ and $\epsilon_H\equiv -\dot{H}/H^2$. After varying the quadratic action with respect to $\delta P$ yields
\begin{align}
    \delta \ddot{P}+3H\delta \dot{P}+\left(M_{PP}+\frac{k^2}{a^2}\right)\delta P+M_{P\phi}\delta\phi+\bar{M}_{P\phi}\delta\dot{\phi}=0,
\end{align}
where we have defined
\begin{align}
    M_{PP}\equiv&-\left(\frac{\ddot{g}}{g}-\frac{H\dot{g}}{g}-2H^2+\epsilon_HH^2\right)+\left(2-\epsilon_H\right)H^2\epsilon_B,\\
    M_{P\phi}\equiv& H^2\sqrt{\epsilon_B}\left[M_{\text{pl}}\epsilon_B\frac{g_{\phi}}{g}-\sqrt{2}\sqrt{\epsilon}\left(\epsilon_H+\eta\right)+\frac{2M_{\text{pl}}}{H}\left(\frac{\dot{g}_{\phi}}{g}-\frac{\dot{g}g_{\phi}}{g^2}\right)\right],\\
    \bar{M}_{P\phi}\equiv& 2M_{\text{pl}}\sqrt{\epsilon_B}H\frac{g_{\phi}}{g}.
\end{align}
When energy density of two-form gauge fields is negligible, i.e., $\epsilon_B=0$, the interactions between scalar-field modes and gauge-field modes $M_{P\phi}=\bar{M}_{P\phi}=0$. And we note the $\epsilon_B$ term in the $M_{PP}$ can be ignored in the slow roll approximation and the sign of the second term is opposite to that of one-form gauge fields, which implies the different unstable conditions of the slow roll inflationary attractors. However, There still exist stable attractors after the destabilization, which can be found in the main text.

Next we treat the vector sector of the models. After gauge-fixing we have one vector perturbation $M_a$ for two-form gauge fields and one perturbation $B_j$ for gravity. After expanding the action around background and performing several integration by parts we have
\begin{align}\label{2vaction2}
    S^{(2)}_{\text{vector}}=\int dtd^3x&\Bigg\{\frac{M_{\text{pl}}^2}{4a}\partial_iB_j\partial_iB_j+\frac{f^2}{a}\Big(\partial_i\dot{\mathbb{W}}_a\partial_i\dot{\mathbb{W}}_a-\frac{\dot{\mathbb{B}}}{2a^2}\partial^2\mathbb{W}_jB_j\Big)\nonumber\\
    &-\frac{f^2}{a^3}\partial_k(\partial_i\mathbb{W}_a)\partial_k(\partial_i\mathbb{W}_a)
\Bigg\}.
\end{align}
To eliminate the non-dynamical vaiable $B_j$, we can use its equation of motion
\begin{equation}
    M_{\text{pl}}^2B_j+\frac{f^2\dot{\mathbb{B}}}{a^2}\mathbb{W}_j=0 \ .
\end{equation}
Substituting it back to the quadratic action, we can deduce the following
\begin{align}\label{2action2v}
    S^{(2)}_{\text{vector}}=\int dtd^3x&\Bigg\{\frac{f^2}{a}\partial_i\dot{\mathbb{W}}_a\partial_i\dot{\mathbb{W}}_a-\frac{f^2}{a^3}\partial_k(\partial_i\mathbb{W}_a)\partial_k(\partial_i\mathbb{W}_a)-\frac{f^4\dot{\mathbb{B}}^2}{2M_{\text{pl}}^2a^5}\partial_i\mathbb{W}_a\partial_i\mathbb{W}_a
    \Bigg\}.
\end{align}

Finally, we treat the tensor perturbations $(w_{ij}, \mathbb{R}_{ai})$, one is dynamical metric mode and the other is non-dynamical gauge-field mode. After expanding the action around background and performing several integration by parts, then also using the background equations of motion, the quadratic action is given by
\begin{align}
    S^{(2)}_{\text{tensor}}=\int dtd^3xa^3&\Bigg\{\frac{M_{\text{pl}}^2}{8}(\dot{w}_{ij})^2-\frac{M_{\text{pl}}^2}{8a^2}(\partial_kw_{ij})^2+\frac{M_{\text{pl}}^2}{4}\frac{\dot{a}}{a}w_{ij}\dot{w}_{ij}
    +\Big(\frac{M_{\text{pl}}^2}{8}\frac{\ddot{a}}{a}+\frac{M_{\text{pl}}^2}{4}\frac{\dot{a}^2}{a^2}\nonumber\\
    &-\frac{\dot{\mathbb{B}}^2}{2a^4}g^2\Big)(w_{ij})^2+\frac{g^2}{2a^4}(\partial_i\mathbb{R}_{aj})^2-\frac{g^2}{a^4}\dot{\mathbb{B}}w_{jk}\epsilon_{kia}\partial_i\mathbb{R}_{aj}
    \Bigg\}.
\end{align}
$\mathbb{R}_{ai}$ appears with no time derivative. After varying this quadratic action with respect to $\mathbb{R}_{aj}$ yields the constraint equation
\begin{equation}
    \partial_i\mathbb{R}_{aj}=\dot{\mathbb{B}}w_{jk}\epsilon_{kia}.
\end{equation}
Then substituting it back to the quadratic action we obtain
\begin{align}
     S^{(2)}_{\text{tensor}}=\int dtd^3xa^3&\Bigg\{\frac{M_{\text{pl}}^2}{8}(\dot{w}_{ij})^2-\frac{M_{\text{pl}}^2}{8a^2}(\partial_kw_{ij})^2+\frac{M_{\text{pl}}^2}{4}\frac{\dot{a}}{a}w_{ij}\dot{w}_{ij}\nonumber\\
    &+\Big(\frac{M_{\text{pl}}^2}{8}\frac{\ddot{a}}{a}+\frac{M_{\text{pl}}^2}{4}\frac{\dot{a}^2}{a^2}-\frac{3\dot{\mathbb{B}}^2}{2a^4}g^2\Big)(w_{ij})^2
\end{align}
where $\mathbb{R}_aj$ has been eliminated so the only dynamical degrees of freedom are metric perturbations.

\section{Slow roll approximation of the quadratic action}\label{2action_sl}
In this appendix we can reduce all of the quadratic actions we derived in the last section in the slow roll approximation.
\subsection{One-form gauge field}
After changing the time variable to the conformal time $d\tau=dt/a$ and introducing the canonical variables (\ref{nv1}), the quadratic action can be written as follows
\begin{align}
    S^{(2)}_{\text{scalar}}=\frac{1}{2}&\int d\tau d^3x \Bigg\{(\Delta_{\phi}')^2-(\partial\Delta_{\phi})^2+(\Delta_{Q}')^2-(\partial\Delta_{Q})^2+(\partial\Delta_{U}')^2-(\partial^2\Delta_{U})^2\nonumber\\
    &+a^2H^2\Big[2-\epsilon_H-\frac{V_{\phi\phi}}{H^2}+2(3+\epsilon_H)\epsilon+4\epsilon\eta+\epsilon_E\Big(-4\sqrt{2\epsilon}\frac{f_{\phi}}{f}-\frac{f_{\phi}^2}{f^2}\nonumber\\
    &+3\frac{f_{\phi\phi}}{f}\Big)\Big](\Delta_{\phi})^2+\Big[\frac{f''}{f}+a^2H^2\epsilon_E(2\epsilon-6)\Big](\Delta_{Q})^2+\frac{f''}{f}(\partial\Delta_{U})^2\nonumber\\
    &-aH\Big[4\sqrt{2}\sqrt{\epsilon_E}\frac{f_{\phi}}{f}\frac{f'}{f}+aH\sqrt{\epsilon_E}\Big(4\sqrt{2}\frac{f_{\phi}}{f}\epsilon_E-4\sqrt{\epsilon}(\epsilon_H+\eta)\Big)\Big]\Delta_{\phi}\Delta_{Q}\nonumber\\
    &-aH\Big[4\sqrt{2}\frac{f_{\phi}}{f}\sqrt{\epsilon_E}\Big]\Delta_{\phi}\Delta_{Q}'
    \Bigg\}  ,
\end{align}
where $\Delta_U$ is fluctuations of the magnetic field that is decoupled with perturbations of scalar fields and electric fields in the action. In fact, it is an isocurvature mode which does not contribute to the curvature perturbations \cite{Gorji:2020vnh}. Hence we can ignore it. Now we consider the slow roll approximation $\epsilon\ll 1$, $\epsilon_H\ll 1$, $\epsilon_E\ll 1$, $|\eta|\ll 1$ and the de Sitter background evolution of spacetime $1/\tau=-aH$. For the stable attractor (\ref{g_attractors}), we have $f\sim \tau^2$ and  hence
\begin{equation}
   \frac{f'}{f}=\frac{2}{\tau},\ \ \ \ \ \ \ \ \frac{f''}{f}=\frac{2}{\tau^2}  .    
   \label{ff}
\end{equation}
 Moreover, from (\ref{geometry}), we also have 
\begin{equation}
\frac{f_{\phi}}{f}=\frac{2}{L_1},\ \ \ \ \ \ \ \ \frac{f_{\phi\phi}}{f}=\frac{4-2L_{1\phi}}{L^2}\simeq \frac{4}{L_1^2},
\end{equation}
where we have used slow roll approximation $|L_{\phi}|=|\dot{L}/\dot{\phi}|=|\dot{L}/(HL)|=|\eta+\epsilon_H|\ll 1$.  Consequently, the quadratic action reduces to (\ref{action2per})

For the vector modes, we also introduce the canonical variables (\ref{nv1v}). Then we can deduce the quadratic action of vector sector as
\begin{align}
    S^{(2)}_{\text{vector}}=\frac{1}{2}\int d\tau d^3x \Bigg\{ \boldsymbol{\Delta}_{a}'\cdot  \boldsymbol{\Delta}_{a}'-\partial_i \boldsymbol{\Delta}_{a}\cdot\partial_i \boldsymbol{\Delta}_{a}+\Big(\frac{f''}{f}-2a^2H^2\epsilon_E\Big)\boldsymbol{\Delta}_{a}\cdot  \boldsymbol{\Delta}_{a}
    \Bigg\}.
\end{align}
In the slow roll approximation where $\epsilon_E \ll 1$ and (\ref{ff}), the action is reduced to (\ref{vqd}).

Similarly, after introducing the canonical variables (\ref{nv1t}), we deduce the tensor part of the quadratic action of gravity and gauge fields (see appendix \ref{quadratic_action})
\begin{align}
    S^{(2)}_{\text{tensor}}=\frac{1}{8}\int d\tau d^3x &\Bigg\{(h_{ij}')^2-(\partial h_{ij})^2+(t_{ij}')^2-(\partial t_{ij})^2+\Big(\frac{a''}{a}+2a^2H^2\epsilon_E\Big)(h_{ij})^2\nonumber\\
    &+\frac{f''}{f}(t_{ij})^2+4aH\sqrt{\epsilon_E}\frac{f'}{f}h_{ij}t_{ij}-4aH\sqrt{\epsilon_E}h_{ij}t_{ij}'
    \Bigg\} .
\end{align}
After ignoring the slow roll parameter $\epsilon_E$ but keeping $\sqrt{\epsilon_E}$ and using $1/\tau=-aH$, then we can reduce the action to (\ref{1ft}).

\subsection{Two-form gauge field}
The two-form case is similar to the one of one-form, where we introduce the canonical variables (\ref{nv2}) and use the conformal time $\tau$ rather than the cosmological one $t$. We can derive the quadratic action of the fluctuations $\Delta_{\phi}$ and $\Delta_P$
\begin{align}
     S^{(2)}_{\text{scalar}}=\frac{1}{2}&\int d\tau d^3x \Bigg\{(\Delta_{\phi}')^2-(\partial\Delta_{\phi})^2+(\Delta_{P}')^2-(\partial\Delta_{P})^2\nonumber\\
     &+a^2H^2\Big[2-\epsilon_H-\frac{V_{\phi\phi}}{H^2}+2(3+\epsilon_H)\epsilon+4\eta\epsilon+\epsilon_B\Big(-2\sqrt{2\epsilon}\frac{g_{\phi}}{g}-5\frac{g_{\phi}^2}{g^2}\nonumber\\
     &+3\frac{g_{\phi\phi}}{g}\Big)\Big](\Delta_{\phi})^2+\Big[\frac{g''}{g}-2aH\frac{g'}{g}+a^2H^2\epsilon_H+a^2H^2(-3+\epsilon_H)\epsilon_B\Big](\Delta_P)^2\nonumber\\
     &-aH\Big[4\frac{g_\phi}{g}\frac{g'}{g}\sqrt{\epsilon_B}-aH\sqrt{\epsilon_B}\Big(2\frac{g_\phi}{g}(2-\epsilon_B)+2\sqrt{2}(\epsilon_H+\eta)\Big)\Big]\Delta_{\phi}\Delta_P\nonumber\\
     &+aH\sqrt{\epsilon_B}\Big[4\frac{g_{\phi}}{g}-2\sqrt{2}\sqrt{\epsilon}\Big]\Delta_{\phi}\Delta_{P}'\Bigg\}.
\end{align}
We are considering slow roll approximation $\epsilon\ll 1$, $\epsilon_H\ll 1$, $\epsilon_B\ll 1$, $\eta\ll 1$ and de Sitter background $1/\tau=-aH$. For obtaining the scale invariant spectrum of two-form gauge fields, we have $g\sim\tau$, which implies
\begin{equation}\label{gg}
    \frac{g'}{g}=\frac{1}{\tau},\ \ \ \ \ \ \ \frac{g''}{g}=0 .
\end{equation}
 From (\ref{geometry}) and assuming the slow rolling of $L_2$, we also have
\begin{equation}
    \frac{g_{\phi}}{g}=\frac{1}{L_2},\ \ \ \ \ \frac{g_{\phi\phi}}{g}=\frac{1}{L_2^2} \ .
\end{equation}
Then the quadratic action reduces to (\ref{2action2per}).

For the vector modes, after defining the canonical variable (\ref{nv2v}) we can write down the quadratic action of these vector perturbations from (\ref{2action2v})
\begin{align}
    S^{(2)}_{\text{vector}}=\frac{1}{2}\int d\tau d^3x &\Bigg\{ \boldsymbol{\Delta}_{a}'\cdot  \boldsymbol{\Delta}_{a}'-\partial_i \boldsymbol{\Delta}_{a}\cdot\partial_i \boldsymbol{\Delta}_{a}+\Big[\frac{g''}{g}-2aH\frac{g'}{g}+a^2H^2(\epsilon_H\nonumber\\
    &-\frac{\epsilon_B}{2})\Big]\boldsymbol{\Delta}_{a}\cdot  \boldsymbol{\Delta}_{a}
    \Bigg\}.
\end{align}
In the slow roll approximation $\epsilon_H,\epsilon_B\ll 1$ this action is reduced to the same one as one-form case (\ref{vqd}).

For the tensor modes, after using the canonical variable (\ref{nv2t}) the quadratic action reads
\begin{align}
    S^{(2)}_{\text{tensor}}=\frac{1}{8}\int d\tau d^3x \Bigg\{(h_{ij}')^2-(\partial h_{ij})^2+\Big(\frac{a''}{a}-12\epsilon_Ba^2H^2\Big)(h_{ij})^2
    \Bigg\}.
\end{align}
In the leading order of the slow roll approximation $\epsilon_B\ll 1$, we can see the quadratic action is the same as that of the single-field inflation.

\section{Classification of attractors}\label{class}
We should carefully classify the strong and weak coupling of attractors for a general choice of $f(\phi)$ and $g(\phi)$. Here we consider the symmetric potential $V(-\phi)=V(\phi)$. There are two kinds of evolution of $\phi$ for symmetric potential, left-to-right evolution ($\dot{\phi}>0$) and right-to-left evolution ($\dot{\phi}<0$) solutions. For the former case, the scale factor grows with the $\phi$. Hence only if $f_{\phi}<0$ the coupling is weak at the beginning. While for the later case, the scale factor decreases with the $\phi$ and only when $f_{\phi}>0$ we are in the weak regime. We discuss these attractors in three scenarios: $L_{1\pm}(\phi)>0$, $L_{1\pm}(\phi)<0$ and $L_{1-}(\phi)>0$, $L_{1+}(\phi)<0$. 
%$Case\ I$: $L_{\pm}(\phi)>0$. Both solutions are right-to-left. If $f_{\phi\phi}>0$, there is no solution such that $L_{+}>0$ hence is excluded. If $f_{\phi\phi}<0$ only if $f_{\phi}<0$ solutions exist. This is the strong coupled regime hence is unfavorable. $Case\ II$: $L_{\pm}(\phi)<0$. Both solutions are left-to-right. There is no solution such that $L_{-}<0$ for $f_{\phi\phi}>0$ and only if $f_{\phi}>0$ solutions exist for $f_{\phi\phi}<0$, hence can also be excluded. $Case\ III$: $L_{-}(\phi)>0$ and $L_{+}(\phi)<0$. All solutions are strong coupling for $f_{\phi\phi}<0$. When $f_{\phi\phi}>0$, $f_{\phi}<0$ is strong coupling for solution $L_{-}(\phi)$ while $f_{\phi}>0$ is strong coupling for solution $L_{+}(\phi)$. 
We find the only weakly coupled solutions are $-\dot{\phi}/H=L_{1-}(\phi)$ when $f_{\phi\phi}>0$, $f_{\phi}>0$, and $-\dot{\phi}/H=L_{1+}(\phi)$ when $f_{\phi\phi}>0$, $f_{\phi}<0$, which are shown in Table 2. 
\begin{table}[ht]
\begin{center}
 \setlength{\tabcolsep}{6.2mm}{
\begin{tabular}{ll|ll|l|l}
\hline
                                          &    & $L_{1-}>0$                                            & $L_{1+}<0$                        & $L_{1\pm}>0$                      & $L_{1\pm}<0$                      \\ \hline
\multicolumn{1}{l|}{\multirow{2}{*}{$f_{\phi\phi}>0$}} & $f_{\phi}>0$ & \multicolumn{1}{l|}{Weak}                       & Strong                  & \multirow{2}{*}{$\times$}      & \multirow{2}{*}{$\times$}      \\ \cline{2-4}
\multicolumn{1}{l|}{}                     & $f_{\phi}<0$ & \multicolumn{1}{l|}{Strong}                  & Weak                       &                         &                         \\ \hline
\multicolumn{1}{l|}{\multirow{2}{*}{$f_{\phi\phi}<0$}} & $f_{\phi}>0$ & \multicolumn{1}{l|}{\multirow{2}{*}{Strong}} & \multirow{2}{*}{Strong} & \multirow{2}{*}{Strong} & \multirow{2}{*}{Strong} \\ \cline{2-2}
\multicolumn{1}{l|}{}                     & $f_{\phi}<0$ & \multicolumn{1}{l|}{}                        &                         &                         &                         \\ \hline
\end{tabular}
}\caption{Different regimes of the one-form attractors of inflation with potential $V(-\phi)=V(\phi)$. Weak: weakly coupling. Strong: strong coupling. $\times$: do not exist.}
\end{center}
\end{table}\label{tableaa}

The two-form case is similar. We should exclude the cases which run from strong coupling at the beginning to weak coupling at the end of inflation. That is, we only consider $g_{\phi}<0$ for left-to-right rolling of $\phi$ and $g_{\phi}>0$ for right-to-left rolling of $\phi$. After dividing the solutions in three scenarios: $L_{1\pm}(\phi)>0$, $L_{1\pm}(\phi)<0$ and $L_{1-}(\phi)>0$, $L_{1+}(\phi)<0$, we pick up all weak-coupling to strong-coupling solutions and show them in Table 3. We see there are more solutions than case of one-form gauge fields that running from weak coupling to strong coupling.
\begin{table}[ht]
\begin{center}
 \setlength{\tabcolsep}{6.2mm}{
\begin{tabular}{ll|ll|ll|ll}
\hline
\multicolumn{2}{l|}{}                                                 & $L_{2-}>0$                     & $L_{2+}<0$    & \multicolumn{2}{l|}{$L_{2\pm}>0$}              & \multicolumn{2}{l}{$L_{2\pm}<0$}              \\ \hline
\multicolumn{1}{l|}{\multirow{2}{*}{$g_{\phi\phi}>0$}} & $g_{\phi}>0$ & \multicolumn{1}{l|}{Weak}      & Strong        & \multicolumn{2}{l|}{\multirow{2}{*}{$\times$}} & \multicolumn{2}{l}{\multirow{2}{*}{$\times$}} \\ \cline{2-4}
\multicolumn{1}{l|}{}                                  & $g_{\phi}<0$ & \multicolumn{1}{l|}{Strong}    & Weak          & \multicolumn{2}{l|}{}                          & \multicolumn{2}{l}{}                          \\ \hline
\multicolumn{1}{l|}{\multirow{2}{*}{$g_{\phi\phi}<0$}} & $g_{\phi}>0$ & \multicolumn{2}{l|}{\multirow{2}{*}{$\times$}} & \multicolumn{2}{l|}{Weak}                      & \multicolumn{2}{l}{$\times$}                  \\ \cline{2-2} \cline{5-8} 
\multicolumn{1}{l|}{}                                  & $g_{\phi}<0$ & \multicolumn{2}{l|}{}                          & \multicolumn{2}{l|}{$\times$}                  & \multicolumn{2}{l}{Weak}                      \\ \hline
\end{tabular}
}\caption{Different regimes of the two-form attractors of inflation with potential $V(-\phi)=V(\phi)$. Weak: weakly coupling. Strong: strong coupling. $\times$: do not exist.}
\end{center}
\end{table}\label{tablebb}


\end{document}